\documentclass[showpacs,aps,prd,reprint,superscriptaddress,nofootinbib,longbibliography]{revtex4-1}

\usepackage[colorlinks=true, pdfstartview=FitV, bookmarks=true, bookmarksnumbered=true, breaklinks]{hyperref}
\usepackage[dvipdfmx]{graphicx}
\usepackage{amsmath,amssymb,bm,color,longtable,mathrsfs,amsfonts}

\definecolor{DeepPink2}{rgb}{0.932,0.07,0.536}
\definecolor{RoyalBlue1}{rgb}{0.284,0.464,1}
\definecolor{SpringGreen3}{rgb}{0,0.804,0.4}
\hypersetup{linkcolor=RoyalBlue1, citecolor=DeepPink2, urlcolor=SpringGreen3}

\begin{document}
\begin{flushright}
\end{flushright}

\title{Heavy baryon spectrum with chiral multiplets of scalar and vector diquarks}

\author{Yonghee~Kim}
\email[]{\it kimu.ryonhi@phys.kyushu-u.ac.jp}
\affiliation{Department of Physics, Kyushu University, Fukuoka 819-0395, Japan}

\author{Yan-Rui Liu}
\email[]{\it yrliu@sdu.edu.cn}
\affiliation{School of Physics, Shandong University, Jinan 250100, China}

\author{Makoto~Oka}
\email[]{\it oka@post.j-parc.jp}
\affiliation{Advanced Science Research Center, Japan Atomic Energy Agency (JAEA), Tokai 319-1195, Japan}
\affiliation{Nishina Center for Accelerator-Based Science, RIKEN, Wako 351-0198, Japan}

\author{Kei~Suzuki}
\email[]{{\it k.suzuki.2010@th.phys.titech.ac.jp}}
\affiliation{Advanced Science Research Center, Japan Atomic Energy Agency (JAEA), Tokai 319-1195, Japan}

\date{\today}

\begin{abstract}
Chiral effective theory of scalar and vector diquarks is formulated according to the linear sigma model. The main application is to describe the ground and excited states of singly heavy baryons with a charm or bottom quark. Applying the potential quark model between the diquark and the heavy quark ($Q=c, b$), we construct a heavy-quark--diquark model. The spectra of the positive- and negative-parity states of $\Lambda_Q$, $\Sigma_Q$, $\Xi^{(')}_Q$ and $\Omega_Q$ are obtained.
The masses and interaction parameters of the effective theory are fixed partly from the lattice QCD data and also from fitting low-lying heavy baryon masses. We find that the negative parity excited states of $\Xi_Q$ (flavor $\bar{\bf 3}$) are different from those of $\Lambda_Q$, because of the inverse hierarchy of the pseudoscalar diquark.On the other hand, $\Sigma_Q, \Xi'_Q$ and $\Omega_Q$ (flavor ${\bf 6}$) baryons have similar spectra. We compare our results of the heavy-quark--diquark model with experimental data as well as the quark model. 
\end{abstract}
\maketitle

\section{Introduction}
Recent observation of multiquark exotic hadron stimulates discussion on various different structures of hadrons.
Multiquark correlations are often manifested as clusters of quarks. 
The simplest cluster is the diquark made of two quarks \cite{GellMann:1964nj, Ida:1966ev, Lichtenberg:1967zz, Lichtenberg:1967, Souza:1967rms, Lichtenberg:1968zz, Carroll:1969ty, Lichtenberg:1981pp, RevModPhys.65.1199, JAFFE20051}. It was suggested that the diquark plays important roles in the baryon spectra and decays. It is also a main component in the color superconducting phase of dense matter.
For the properties of diquarks, such as the mass and size, lattice QCD simulations have been used. \cite{Hess:1998sd, Orginos:2005vr, Alexandrou:2006cq, Babich:2007ah, DeGrand:2007vu, Green:2010vc, ChinQCD, Watanabe:2021oyv}.

In studying the properties of the diquark, singly heavy baryons ($Qqq$) are useful. They consist of one heavy (charm or bottom) quark ($Q=c, b$) and two light (up, down or strange) quarks ($q=u, d, s$), which can be represented by a diquark ($qq$) characterized by $SU(3)$ \cite{Lichtenberg:1975ap, Lichtenberg:1982jp, Lichtenberg:1982nm, Fleck:1988vm, Ebert:1995fp, Ebert:2007nw, Kim:2011ut, Ebert:2011kk, Chen:2014nyo, Jido:2016yuv, Kumakawa:2017ffl, CETdiquark, scalar, Dmitrasinovic:2020wye, Kawakami:2020sxd, Suenaga:2021qri}. Here the $P$-wave excited states of a singly heavy baryon are classified into two kinds of orbital excitations, the $\lambda$-mode and the $\rho$-mode. The $\lambda$-mode is an excitation of the coordinate between the heavy quark and the diquark. On the other hand, the $\rho$-mode is an excitation of the diquark cluster \cite{Copley:1979wj, yoshida}. 

The most popular light diquarks for hadron spectroscopy is the scalar diquark with the spin-parity $J^P=0^+$, color $\bar{\bf 3}$ and flavor $\bar{\bf 3}$. It appears most frequently in hadrons such as the ground states of $\Lambda_Q$ and $\Xi_Q$.
In Ref.~\cite{CETdiquark}, the authors have applied the chiral effective theory based on $SU(3)_R \times SU(3)_L$ chiral symmetry for this diquark.
There the scalar and pseudoscalar diquarks are chiral partners to each other, {\it i.e.}, belonging to the same representation of chiral symmetry. Their mass difference comes from spontaneous chiral symmetry breaking, and they become degenerate when chiral symmetry is recovered. This theory also shows that $U_A(1)$ anomaly~\cite{tHooft:1976snw, tHooft:1986ooh} leads to inverse hierarchy of diquarks masses, that is, the non-strange pseudoscalar diquark is heavier than that containing one strange quark, $M(ud, 0^-) > M(ds/su, 0^-)$. 
As a consequence, the diquark cluster model of charmed baryons predicts that the $\rho$-mode excited states are also reversely ordered as $M_\rho(\Lambda_{c}, 1/2^-) > M_\rho(\Xi_{c}, 1/2^-)$.

In the previous work, we calculated the spectrum of $\Lambda_Q$ and $\Xi_Q$ baryons as two body systems of a  spin $0$ diquark and a heavy quark \cite{scalar}.
In this paper, we extend our approach to include spin-1 ($1^+$ and $1^-$) diquarks and consider the flavor sextet baryons, $\Sigma_Q$, $\Xi'_Q$ and $\Omega_Q$.  We present a chiral effective theory of the scalar/pseudoscalar ($0^\pm$) and vector/axial-vector ($1^\pm$) diquarks with $SU(3)_R \times SU(3)_L$ symmetry.
We also construct the heavy-quark--diquark model for the flavor sextet baryons, $\Sigma_Q$, $\Xi'_Q$ and $\Omega_Q$.
The parameters of the effective Lagrangian and the potential of the heavy-quark--diquark model are determined so that they reproduce the observed baryon masses.
We find that the spectra of $\Lambda_Q$ and $\Xi_Q$ contain the $\lambda$-mode of the scalar diquark, $\rho$-mode from the pseudoscalar diquark and also from the vector diquark. We compare the results with experimental data and also the quark model.

This paper is organized as follows.
In Sec.~\ref{Sec_2}, we introduce diquarks and formulate chiral effective theory of diquarks.
In Sec.~\ref{Sec_3}, we construct the heavy-quark--diquark potential model and determine parameters of the potential between the heavy quark and the diquark.
In Sec.~\ref{Sec_4}, we show numerical results of the singly heavy baryon spectrum.
At last, Sec.~\ref{Sec_5} is devoted to our conclusion and outlook.

\section{Chiral effective theory of diquark} \label{Sec_2}

In this section, we introduce a chiral effective theory of diquarks and mass formulas of diquarks given by spontaneous chiral symmetry breaking, both for the scalar and pseudoscalar diquarks \cite{CETdiquark, scalar}, and for the vector and axial-vector diquarks.  
\subsection{Scalar and pseudoscalar diquarks}
We here consider diquarks belonging to the color $\bf \overline{3}$ representation given in Table \ref{diop}.
They are the diquarks appearing in the baryon, color-singlet three-quark system.
 \begin{table}[bt]
  \centering
    \caption{Local diquark operators in color $\bf \overline{3}$.
     $C=i \gamma^0 \gamma^2$ is the charge conjugation operator. 
     The subscripts $S$ and $A$ indicate the flavor symmetric ($S$, {\bf 6}) and antisymmetric ($A$, $\bar{\bf 3}$)  operators, respectively. }      
  \begin{tabular}{ l | c | c | c | c  } \hline\hline
    Diquark & Operator & $J^P$ & Color & Flavor   \\ \hline \hline
 scalar & $(q^T C\gamma^5 q)^{\overline{3}}_{A}$ &$0^+$&$\overline{3}$&$\overline{3}$\\ 
 pseudoscalar& $(q^T Cq)^{\overline{3}}_{A}$ &
                                                                                    $0^-$&$\overline{3}$&$\overline{3}$\\ 
 vector & $(q^T C\gamma^{\mu} \gamma^5 q)^{\overline{3}}_{A}$ &
                                                                                         $1^-$&$\overline{3}$&$\overline{3}$\\    
 axial-vector & $(q^T C\gamma^{\mu} q)^{\overline{3}}_{S}$ &
                                                                                         $1^+$&$\overline{3}$& 6 \\ \hline \hline
  \end{tabular}  
  \label{diop}
  \end{table}

To consider the diquark from the viewpoints of chiral symmetry, we use the chiral projection operators $P_{R,L}\equiv (1 \pm \gamma^5)/2,~\gamma^{5}=i\gamma^{0}\gamma^{1}\gamma^{2}\gamma^{3}$. We split the quark operator $q^a_i$ into the right quark as $q^a_{R,i}=P_{R}q^a_i$ and the left quark as $q^a_{L,i}=P_{L}q^a_i$, where $a$ and $i$ are color and flavor indices of the quark, respectively. 
Under the chiral $SU(3)_R \times SU(3)_L$ transform given by 
\begin{eqnarray}
\begin{split}
&q^a_{R,i} \rightarrow (U_R)_{ij} q^a_{R,j},~~U_R \in SU(3)_R,~~~~~(3,1)\\
&q^a_{L,i} \rightarrow (U_L)_{ij} q^a_{L,j},~~U_L \in SU(3)_L,~~~~~(1,3)
\end{split}
\label{rlqt}
\end{eqnarray}
the diquarks of spin 0 and 1 transform as $(\bar{3}, 1)$, $(1,\bar 3)$ and $(3,3)$ representations, respectively, as shown in Table \ref{diopchiral}.

 \begin{table}[b]
  \centering
    \caption{Diquark operators in the chiral basis.}   
  \begin{tabular}{ l | c | c | c  } \hline\hline
   Chiral operator & Spin & Color & Chiral \\ \hline \hline
 $d^a_{R,i} = \epsilon_{abc}\epsilon_{ijk}(q^{bT}_{R,j} Cq^c_{R,k})$ &
 										$0$&$\overline{3}$&$(\overline{3}, 1)$\\ 
 $d^a_{L,i} = \epsilon_{abc}\epsilon_{ijk}(q^{bT}_{L,j} Cq^c_{L,k})$ &
                                                                                    $0$&$\overline{3}$&$(1, \overline{3})$\\ 
 $d^{a,\mu}_{ij} = \epsilon_{abc}(q^{bT}_{L,i} C \gamma^\mu q^c_{R,j})$ &
                                                                                         $1$&$\overline{3}$&$(3,3)$\\  \hline \hline
  \end{tabular}  
  \label{diopchiral}
  \end{table}

Chiral effective theory for the scalar ($0^+$) and pseudoscalar ($0^-$) diquarks with color $\bf \overline{3}$ and flavor $\bf \overline{3}$ is given according to Ref. \cite{CETdiquark}, by 
\begin{eqnarray}
\mathcal{L}=\mathcal{L}_S + \frac{1}{4}{\rm Tr}[\partial^\mu \Sigma^\dag \partial_\mu \Sigma]-V(\Sigma),
\label{elagsub}
\end{eqnarray}
\begin{eqnarray}
\begin{split}
\mathcal{L}_S&=\mathcal{D}_\mu d_{R,i}(\mathcal{D}^\mu d_{R,i})^\dag + \mathcal{D}_\mu d_{L,i}(\mathcal{D}^\mu d_{L,i})^\dag \\
		   &-m_{S0}^2(d_{R,i}d_{R,i}^\dag+d_{L,i}d_{L,i}^\dag) \\
		   &-\frac{m_{S1}^2}{f_\pi}(d_{R,i}\Sigma_{ij}^\dag d_{L,j}^\dag+d_{L,i}\Sigma_{ij}d_{R,j}^\dag) \\
		   &-\frac{m_{S2}^2}{2f_\pi^2}\epsilon_{ijk}\epsilon_{lmn} (d_{R,k}\Sigma_{li}\Sigma_{mj}d_{L,n}^\dag+d_{L,k}\Sigma^\dag_{li}\Sigma^\dag_{m,j}d_{R,n}^\dag). \\
\end{split}
\label{slag}
\end{eqnarray}

Here $\Sigma$ represents meson field of chiral $(\bar{3},3)$ representation \cite{GellMann:1960np, levy1967currents}. 
This contains nonet scalar $\sigma$ and pseudoscalar $\pi$ mesons,
 whose chiral transform is given by
\begin{equation}
\Sigma_{ij}=\sigma_{ij}+i\pi_{ij} \rightarrow U_{L,ik} \Sigma_{km} U^{\dag}_{R,mj},~~~(\overline{3},3)
\end{equation}
In Eq.(\ref{elagsub}), $V(\Sigma)$ is the potential for the meson field $\Sigma$ and
is supposed to induce spontaneous chiral symmetry breaking  (SCSB), represented by the vacuum expectation values of the scalar meson $\sigma$ as   
\begin{equation}
\langle \Sigma_{ij} \rangle =\langle \sigma_{ij} \rangle =f_{\pi} \delta_{ij}~,~~~\langle \pi_{ij} \rangle =0,
\label{vevmeson}
\end{equation}
where $f_\pi \simeq 92$ MeV is the pion decay constant. Then $\pi$ is regarded as the massless Nambu-Goldstone bosons.
Also the parity transformation is given as $\mathcal{P}~:~\Sigma \rightarrow \Sigma^{\dag}$.

Since the diquark is not a color-singlet state, 
we introduce the color-gauge-covariant derivative, $\mathcal{D}^{\mu}=\partial^{\mu}+igT^{\alpha}G^{\alpha,\mu}$, with the gluon field $G^{\mu}$ and the color $SU(3)$ generator for the color $\bf \bar{3}$ representation  $T^{\alpha}$. All the color indices are contracted and not explicitly written.

 For each term in Eq.~(\ref{slag}), the parameters $m_{S0}^2, m_{S1}^2$, and $m_{S2}^2$ give the masses of scalar and pseudoscalar diquarks. When the chiral symmetry is restored, the scalar and pseudoscalar diquarks are degenerate with the chiral invariant mass $m_{S0}$. In the ordinary vacuum, the mass splitting is induced by SCSB and is expressed by $m_{S1}$ and $m_{S2}$. Besides, the $m_{S1}^2$ term brings about the $U_A(1)$ symmetry breaking~\cite{tHooft:1976snw, tHooft:1986ooh}.

When the flavor $SU(3)$ symmetry is broken, the masses of scalar and pseudoscalar diquarks, $M(0^+)$ and $M(0^-)$, satisfy the relation given by \cite{CETdiquark}
\begin{eqnarray}
[M_{qs}(0^+)]^2 - [M_{qq}(0^+)]^2 = [M_{qq}(0^-)]^2 - [M_{qs}(&&0^-)]^2 \nonumber \\
												&&> 0.
\label{inverse}
\end{eqnarray}
The index $q$ stands for $u$ or $d$ quark and $s$ for the $s$ quark, which means the constituent quark of diquarks.
Here we can find the inverse mass hierarchy for the pseudoscalar diquarks, where the non-strange diquark is heavier than the singly strange diquark, $M_{qq}(0^-) > M_{qs}(0^-)$.

\subsection{Chiral effective Lagrangian including vector and axial-vector diquarks}
Next we introduce the vector ($1^-$) and axial-vector ($1^+$) diquarks, which are both color $\bf \overline{3}$ but have different flavor with each other. 
The chiral diquarks are given in the third line of Table \ref{diopchiral},
\begin{equation}
d^{a,\mu}_{ij} = \epsilon^{abc}(q^{bT}_{L,i}C\gamma^{\mu}q^{c}_{R,j})
                       =\epsilon^{abc}(q^{bT}_{R,j}C\gamma^{\mu}q^{c}_{L,i}).
\label{cvdo}                          
\end{equation}
This contains both right and left quark operators, belonging to chiral $(3,3)$ representation. 
According to Eq. (\ref{rlqt}), the chiral transformation is given as
\begin{equation}
\begin{split}
&d^{a,\mu}_{ij} \rightarrow U_{L,i k} ~d^{a,\mu}_{km} ~U^T_{R,m j},~~~(3,3)\\
&d^{a,\mu \dag}_{ij} \rightarrow U^{T \dag}_{R,i k} ~d^{a,\mu \dag}_{k m} ~U^{\dag}_{L,m j},~~~(\overline{3},\overline{3})\\
\end{split}
\label{cvdt}
\end{equation}

As the spatial inversion interchanges the left and right diquarks, the vector diquarks transform as
\begin{equation}
\mathcal{P}~:~d^{a,\mu}_{ij} \rightarrow -d^{a}_{\mu,ji},
\end{equation}
and then obtain the Lorentz vector and axial-vector diquarks, as
\begin{eqnarray}
&& V^{a,\mu}_{ij}=\frac{1}{\sqrt{2}}(d^{a,\mu}_{ij}-d^{a,\mu}_{ji})
		     =\epsilon_{abc}\frac{1}{\sqrt{2}}(q_i^{bT}C\gamma^{\mu}\gamma^{5}q_j^c), \label{Vfield} \\
&& A^{a,\mu}_{ij}=\frac{1}{\sqrt{2}}(d^{a,\mu}_{ij}+d^{a,\mu}_{ji})
		     =\epsilon_{abc}\frac{1}{\sqrt{2}}(q_i^{bT}C\gamma^{\mu}q_j^c). \label{Afield}
\end{eqnarray}
Thus we see that vector and axial-vector diquarks are chiral partners, belonging to $(3,3)$ representation of chiral symmetry, satisfying $V^{a,\mu}_{ij}=-V^{a,\mu}_{ji}$, $A^{a,\mu}_{ij}=A^{a,\mu}_{ji}$.
Note that the vector diquark belongs to flavor antisymmetric $\bf \overline{3}$ representation, while the axial-vector diquark to symmetric $\bf 6$ representation. 

We now construct the full effective Lagrangian with both the scalar and vector diquarks as
\begin{eqnarray}
&&\mathcal{L}=\mathcal{L}_S + \mathcal{L}_V + \frac{1}{4}{\rm Tr}[\partial^\mu \Sigma^\dag \partial_\mu \Sigma]-V(\Sigma)+\mathcal{L}_{V-S}, 
\label{elag}
\end{eqnarray}
in which the effective Lagrangian for vector and axial-vector diquarks is expressed as
\begin{eqnarray}
\begin{split}
\mathcal{L}_V&=\frac{1}{2}{\rm Tr}[F^{\mu\nu}F^{\dag}_{\mu\nu}]
                    +m_{V0}^2{\rm Tr}[d^{\mu}d^{\dag}_{\mu}] \\
                    &+\frac{m_{V1}^2}{f_{\pi}^{2}}{\rm Tr}[\Sigma^{\dag}d^{\mu}\Sigma^Td^{\dag T}_{\mu}] \\
                    &+\frac{m_{V2}^2}{f_{\pi}^{2}}[{\rm Tr}\{ \Sigma^{T}\Sigma^{\dag T} d_{\mu}^{\dag}d^{\mu} \}
                    					     +{\rm Tr}\{ \Sigma\Sigma^{\dag} d^{\mu}d^{\dag}_{\mu} \}]. \\
\end{split}
\label{vlag}
\end{eqnarray}

For the first term of Eq.~(\ref{vlag}), $F^{\mu \nu}=\mathcal{D}^{\mu}d^{\nu}-\mathcal{D}^{\nu}d^{\mu}$ is the strength of chiral vector diquark fields. 
Similarly to the scalar and pseudoscalar diquarks, masses of vector and axial-vector diquarks are given by three parameters $m_{V0}^2 ,m_{V1}^2$ and $m_{V2}^2$. $m_{V0}$ is the chiral invariant mass of vector and axial-vector diquarks, and their mass splitting is given by the $m_{V1}^2$ and $m_{V2}^2$ terms.  

We neglect terms that includes ${\rm Tr}[\Sigma \Sigma^{\dag}]$, or more than two $\Sigma$'s. 
All the color indices are contracted and the self-interaction terms of gluons are omitted as same as Eq.~(\ref{slag}).

The last term of the effective Lagrangian $\mathcal{L}_{V-S}$ describes the coupling of scalar/vector diquarks and the meson. We neglect this term since it does not contribute to the mass formulas of diquarks.

\subsection{Masses of vector and axial-vector diquarks}

Chiral symmetry is exact if the light quark masses are zero.
In this limit, the effective quark masses of quarks, of oder 300--500 MeV, are given in the linear sigma model by the condensates of the scalar meson $\sigma$, as
$m_{\rm eff}=g_s\langle\sigma\rangle$, 
where $g_s\simeq 3$ is the quark-meson coupling constant.
In reality, chiral symmetry is explicitly broken by the non-zero quark masses.
In order to include this effect, we superpose the bare quark mass and the scalar condensate as

\begin{eqnarray}
&&\mathcal{M}_{\rm eff}=\mathcal{M}+g_s\langle\Sigma\rangle ,
\label{MM}
\end{eqnarray}
where $\mathcal{M}={\rm diag}(m_u, m_d, m_s)$ is the current quark mass matrix. 
The vacuum expectation value $\langle \Sigma \rangle$ may also contain the symmetry breaking with 
$\langle \sigma_{11} \rangle = \langle \sigma_{22} \rangle = f_\pi \simeq 92$ MeV for the $u, d$ quarks and 
$\langle \sigma_{33} \rangle = f_s = 2f_K -f_\pi \simeq 128$ MeV for the $s$ quark.  
By neglecting $u$ and $d$ quark masses, we obtain the effective quark mass matrix as 
\begin{eqnarray}
&&\mathcal{M}_{\rm eff} \simeq g_s f_\pi{\rm diag}(1,1,1+\epsilon) \equiv g_s f_\pi X, \\
&&\epsilon = \frac{f_s}{f_\pi} \left( 1+\frac{m_s-g_sf_\pi}{g_s f_s} \right) \simeq 2/3.
\end{eqnarray}

 We now introduce the explicit chiral symmetry breaking
to the diquark effective Lagrangian by the replacement \cite{CETdiquark}
 \begin{eqnarray}
 \Sigma \rightarrow\tilde{\Sigma} \equiv \Sigma + \mathcal{M}/g_s,
 \label{Sigmatilde}
\end{eqnarray}
with its vacuum expectation value, 
\begin{eqnarray}
 \langle \tilde{\Sigma} \rangle = \mathcal{M}_{\rm eff}/g_s=f_\pi X.
\end{eqnarray}
By applying Eq.~(\ref{Sigmatilde}) to Eq.~(\ref{vlag}) and using the notations of Eqs.~(\ref{Vfield}) and (\ref{Afield}), 
we can read the mass terms for the vector and axial-vector diquarks as 
\begin{eqnarray}
\mathcal{L}_{Vmass}
&& =\frac{1}{2}m_{V0}^2{\rm Tr}[A^{\mu}A^{\dag}_{\mu}]
+\frac{1}{2}m_{V0}^2{\rm Tr}[V^{\mu}V^{\dag}_{\mu}] \nonumber \\
   && +\frac{1}{2}m_{V1}^2{\rm Tr}[XA^{\mu}XA^{\dag}_{\mu}]
   -\frac{1}{2}m_{V1}^2{\rm Tr}[XV^{\mu}XV^{\dag}_{\mu}] \nonumber \\
   && +m_{V2}^2{\rm Tr}[X^2 A^{\mu}A^{\dag}_{\mu}]  
   +m_{V2}^2{\rm Tr}[X^2 V^{\mu}V^{\dag}_{\mu}].
\label{vlagmass}
\end{eqnarray}
Then the masses of axial-vector and vector diquarks, $M(1^+)$ and $M(1^-)$, are given 
to the leading order in $\epsilon$ by
\begin{eqnarray}
 \left[M_{qq} (1^+)\right]^2= m_{V0}^2 + m_{V1}^2 + 2m_{V2}^2, \label{mass1} \\
 \left[M_{qs} (1^+)\right]^2= m_{V0}^2 + m_{V1}^2 + 2m_{V2}^2 \nonumber \\
 +\epsilon(m_{V1}^2+2m_{V2}^2), \label{mass2} \\
 \left[M_{ss} (1^+)\right]^2= m_{V0}^2 + m_{V1}^2 + 2m_{V2}^2 \nonumber \\
  +2\epsilon(m_{V1}^2+2m_{V2}^2), \label{mass3} \\
 \left[M_{qq} (1^-)\right]^2= m_{V0}^2 - m_{V1}^2 + 2m_{V2}^2, \label{mass4} \\
 \left[M_{qs} (1^-)\right]^2= m_{V0}^2 -m_{V1}^2 + 2m_{V2}^2 \nonumber \\
  +\epsilon(-m_{V1}^2+2m_{V2}^2). \label{mass5}
\end{eqnarray}
From Eqs.~(\ref{mass1})--(\ref{mass3}), we find the generalized Gell-Mann--Okubo mass formula for the axial-vector diquarks, \cite{gomass1,gomass2}
\begin{eqnarray}
 [M_{ss} (1^+)]^2 -  [M_{qs} (1^+)]^2 &&=  [M_{qs} (1^+)]^2 -  [M_{qq} (1^+)]^2 \nonumber\\
&&= \epsilon(m_{V1}^2+2m_{V2}^2). 
\label{axial-sqmassdiff}
\end{eqnarray} 
This equation indicates the inequality, 
\begin{eqnarray}
M_{ss}(1^+)+M_{qq}(1^+) < 2M_{qs}(1^+) \label{in}
\end{eqnarray}
One can check this relation from the masses of the singly heavy baryons.
From the spin-averaged masses, $M(\Sigma_c, 1/2^+, 3/2^+)=2496.6$ MeV, 
$M(\Xi'_c, 1/2^+, 3/2^+)=2623.5$ MeV, and $M(\Omega_c, 1/2^+, 3/2^+)=2742.3$ MeV \cite{PDG},
we find the average of the $\Sigma_c$ and $\Omega_c$ masses is smaller than
the $\Xi'_c$ mass as is expected from the inequality (\ref{in}). 

{On the other hand, from Eqs.~(\ref{mass4}) and (\ref{mass5}), the square mass difference between these vector diquarks is given by
\begin{eqnarray}
 [M_{qs} (1^-)]^2 -  [M_{qq} (1^-)]^2 = \epsilon(-m_{V1}^2+2m_{V2}^2). 
\label{vec-sqmassdiff}
\end{eqnarray}
Comparing with Eq.~(\ref{axial-sqmassdiff}), the mass difference by the number of constituent strange quarks is different between the vector and axial-vector diquarks, which is determined by the parameter $m_{V1}^2$. 

\section{Heavy baryon spectrum} \label{Sec_3}

In this section, we construct a nonrelativistic potential model for singly heavy baryons, composed of a heavy quark and a diquark (heavy-quark--diquark model). We determine the unknown model parameters such as the diquark masses and the coefficients of heavy-quark--diquark potential from experimental data of heavy baryons. We refer to a general analysis of the heavy baryon spectrum to Ref.~\cite{Chen:2016spr}. 

\subsection{Heavy-quark--diquark model} \label{Sec_3-1}
We apply a nonrelativistic potential model to the system consisting of a heavy quark and a diquark. 
In Table \ref{tabledb}, we summarize the baryon states that we consider in the present calculation.
We consider the S (scalar), P (pseudoscalar), V (vector) and A (axial-vector) diquarks.
Excited states are labeled by the number of nodes in the radial wave function, $n$, and the relative angular momentum, $L$.

In the quark model, the $P$-wave excited states of baryons are often classified into the
$\lambda$ and $\rho$ modes.
This is based on the quark model notation that denotes the relative coordinate between the heavy quark and the diquark as $\lambda$, while the internal coordinate of the diquark is called $\rho$.
In the heavy-quark--diquark model, the $\lambda$ modes are given by the excitations between the heavy quark and the diquark.
On the contrary, the $\rho$ coordinate is not explicitly treated dynamically, and 
the $\rho$ mode excitations are realized by the excited diquarks, such as P and V diquarks. 
We further consider the nodal excitations in the relative S wave states.
Possible spin-parity $J^{P}$ combinations of the baryon are listed in the last column of Table \ref{tabledb}.
All these states will appear in the final calculation.
\begin{table}[tb]
  \centering
    \caption{Quantum numbers of the ground and excited states of the singly heavy baryons in the 
    heavy-quark--diquark model. }
  \begin{tabular}{r | c | r | r | l} \hline\hline
    & Diquark ($J^P$) &  \quad $n$ &  \quad$L$ & \quad $J^P$ \\ 
      \hline\hline
    			& 			 	&0 &0 	&\quad $1/2^+$   \\ \cline{3-5}
  			& S ($0^+$) 		&0 &1	&\quad $1/2^-, 3/2^-$   \\ \cline{3-5}
 $\Lambda_{c/b}$ 	&     	&0 &2	&\quad $3/2^+, 5/2^+$   \\ \cline{3-5}
 $\Xi_{c/b}$		& 	 		&1 &0	&\quad $1/2^+$   \\ \cline{2-5}
 		& P ($0^-$)          		&0&0 	&\quad $1/2^-$    \\ \cline{2-5}
    		& V ($1^-$)	 		&0&0	&\quad $1/2^-, 3/2^-$     \\  \hline   
    $\Xi'_{c/b}$	& 			&0&0	&\quad $1/2^+, 3/2^+$    \\ \cline{3-5}
    $\Sigma_{c/b}$	& A ($1^+$)	&0&1	&\quad $(1/2^-)^2, (3/2^-)^2, 5/2^-$            \\ \cline{3-5}
    $\Omega_{c/b}$& 			&0&2	&\quad $1/2^+, (3/2^+)^2, (5/2^+)^2, 7/2^+$  \\\cline{3-5}
                         & 		 	&1&0 	&\quad $1/2^+, 3/2^+$   \\ \hline 
\hline
  \end{tabular}
  \label{tabledb}
\end{table}

Following the previous study~\cite{scalar}, we employ the nonrelativistic Hamiltonian for the heavy-quark--diquark model given by
\begin{equation}
H=\frac{{\bm p}^2}{2\mu}+M_Q+M_d+V_{0}+V_{S}, 
\label{ham}
\end{equation}
and the relative momentum and the reduced mass are defined by
\begin{eqnarray}
&& {\bm p}=\frac{M_Q{\bm p}_d-M_d{\bm p}_Q}{M_Q+M_d},\\
&& \mu=\frac{M_Q M_d}{M_Q+M_d}. \label{reduced mass}
\end{eqnarray}
where the indices $Q$ and $d$ denote the heavy quark and the diquark, respectively. 
Also $\bm{p}_{Q/d}$ and $M_{Q/d}$ are the momentum and mass, respectively. 
We solve the Schr$\ddot{\rm o}$dinger equation for this Hamiltonian, applying the Gaussian expansion method \cite{GEM1,GEM2}.

The two-body potential is given in terms of the relative coordinate 
$r=|{\bm r}_d - {\bm r}_Q|$ and consists of the spin-independent central force, $V_{0}(r)$ and
the spin-dependent term, $V_{S}(r)$.
The spin-independent potential $V_{0}(r)$ consists of the Coulomb term with the coefficient $\alpha$, the linear term with $\lambda$ and a constant shift $C$, 
\begin{equation}
V_{0}(r) = -\frac{\alpha}{r}+\lambda r+C .
\label{spin-independent-potential}
\end{equation}
We choose three different choices of the parameters taken from the quark models
by Yoshida et al.~(Y) \cite{yoshida}, Silvestre-Brac (S) \cite{silvestre}, 
and Barnes et al. (B) \cite{barnes}.

$V_{S}$ is the sum of the spin-spin potential $V_{ss}(r)$, the spin-orbit potential $V_{so}(r)$, and the 
tensor potential $V_{ten}(r)$, taken from Refs.~\cite{yoshida, silvestre, barnes}.
The spin-spin potential for the B-potential is taken as a Gaussian function,
\begin{eqnarray}
V_{ss}(r)=({\bm s}_d \cdot {\bm s}_Q)\frac{\kappa_Q}{M_d M_Q} \left( \frac{\Lambda}{\sqrt{\pi}} \right)^3 \exp{(-\Lambda^2 r^2)}.
\label{ssb}
\end{eqnarray}
On the other hand, the Y- and S-potentials have the Yukawa form as
\begin{eqnarray}
V_{ss}(r)=({\bm s}_d \cdot {\bm s}_Q)\frac{\kappa_Q}{M_d M_Q} \frac{\Lambda^2}{r} \exp{(-\Lambda r)}.
\label{ssys}
\end{eqnarray}
Here $\Lambda$ is a cutoff parameter and $({\bm s}_d \cdot {\bm s}_Q)$ is the inner product of the diquark spin ${\bm s}_d$ and the heavy-quark spin ${\bm s}_Q$.
The coefficient $\kappa_Q$ will be determined for the charm and bottom sectors separately in the heavy-quark--diquark model.

The spin--orbit potential $V_{so}(r)$ consists of the symmetric ($+$) and antisymmetric ($-$) terms as
\begin{eqnarray}
V_{so}(r) &&= V_{so}^+(r) +V_{so}^-(r)\label{sopot}\\
 V_{so}^+(r) &&= \eta \frac{(1-e^{-\Lambda r})^2}{r^3} \nonumber \\
  && \times \left(\frac{1}{M_d^2}+ \frac{1}{M_Q^2} +\frac{4}{M_d M_Q} \right)( {\bm L} 
  \cdot {\bm S}^+) ,\label{soplus} \\
   V_{so}^-(r) &&= \eta \frac{(1-e^{-\Lambda r})^2}{r^3} 
   \left( \frac{1}{M_d^2}-\frac{1}{M_Q^2} \right) 
  ( {\bm L}\cdot {\bm S}^-) ,
\label{sominus}
\end{eqnarray}
with ${\bm S}^{\pm} \equiv {\bm s}_d \pm {\bm s}_Q$.
The tensor potential $V_{ten}(r)$ is given by
\begin{eqnarray}
V_{ten}(r) &&= 2\eta \frac{(1-e^{-\Lambda r})^2}{M_d M_Q r^3} \nonumber 
\\
&& \times \left[\frac{3({\bm s}_d \cdot {\bm r})( {\bm s}_Q \cdot {\bm r})}
{r^2} - {\bm s}_d\cdot {\bm s}_Q \right]
\label{tenpot}
\end{eqnarray}
We note that the spin--orbit and tensor potentials were originally included only in the Y-potential 
but we apply the same form also for the S- and B-potentials.
In Eqs.~(\ref{soplus})--(\ref{tenpot}), $\eta$ is the common coupling strength that is
determined later for the heavy-quark--diquark model.

\subsection{Diquark masses and potential parameters}

Choices of the heavy-quark--diquark model parameters are summarized in Table \ref{paraall}.
Among them, the heavy-quark masses, $M_Q$, the strength of the Coulomb potential, $\alpha$, that of the linear confinement potential, $\lambda$, and the cutoff parameter, $\Lambda$, are taken from the quark models, Ref.~\cite{yoshida,silvestre,barnes}.
The constant shifts $C_Q$ in $V_{0}(r)$ are determined by fitting the masses of the ground $\Lambda_Q$ baryon as in Ref.~\cite{scalar}.
\begin{table}[tb]
  \centering
    \caption{Parameters of the heavy-quark--diquark model. 
The values with asterisk (*) are taken from the potential models by Yoshida et al.~(Y-pot.) \cite{yoshida}, by Silvestre-Brac (S-pot.) \cite{silvestre}, and by Barnes et al.~(B-pot.) \cite{barnes}. $\mu$ included in the parameter of the Y-pot. is the reduced mass, Eq.~(\ref{reduced mass}).} 
  \begin{tabular}{ l  | r | r | r  } \hline\hline 
        \multicolumn{1}{ l | }{} & \multicolumn{3}{ c }{Potential models} \\ \cline{2 - 4} 
  Parameters  & ~Y-pot.~\cite{yoshida} & ~S-pot.~\cite{silvestre} & ~B-pot.~\cite{barnes}   \\ \hline \hline
  $\alpha$  			& 60/$\mu$*  & $0.5069$* & $0.7281$*   \\ 
  $\lambda ({\rm GeV}^2)$ & $0.165$*        & $0.1653$* & $0.1425$*    \\
  $ C_c (\rm GeV)$            & $-0.831$  &$-0.707$& $-0.191$ \\
  $ C_b (\rm GeV)$            & $-0.819$  &$-0.696$&  $\ldots$ \\
  $\Lambda (\rm GeV)$     & $0.691$*  &$0.434$* & $1.0946$* \\
  $\kappa_c$                     & $0.8586$&$1.1570$& $2.7258$ \\
  $\kappa_b$                     & $0.6635$&$0.8145$& $\ldots$ \\
 $\eta$				& $0.2494$&$0.2494$&$0.2494$ \\ 
  $ M_c (\rm GeV)$             & $1.750$*  &$1.836$*&  $1.4794$* \\
  $ M_b (\rm GeV)$            & $5.112$*  &$5.227$*&  $\ldots$ \\ \hline\hline
  \end{tabular}
  \label{paraall}
\end{table}

We omit the bottom part of the B-potential parameters, because the original B-potential was not applied to the bottom baryons.

In Table \ref{diquarkmass}, the diquark masses and the parameters of the effective Lagrangian are summarized.
The values for the scalar and pseudoscalar diquarks are taken from Ref.~\cite{scalar}. 
In particular, the mass of the $ud$ scalar diquark, $M_{qq}(0^+)$, is the input value from lattice QCD simulations~\cite{ChinQCD}.
The vector diquark masses, $M_{qq}(1^+), M_{qs}(1^+)$, and $M_{qq}(1^-)$, 
are determined from some spin-averaged charmed baryon masses as inputs:
\begin{eqnarray}
&& M(\Sigma_c,1/2^+,3/2^+)=2496.60 \hbox{ MeV},\nonumber\\
&& M(\Xi'_c,1/2^+,3/2^+)=2623.52 \hbox{ MeV},\nonumber\\
&& M_\rho(\Lambda_c,1/2^-,3/2^-)=2927.65 \hbox{ MeV},\nonumber
\end{eqnarray}
where $M_\rho(\Lambda_c,1/2^-,3/2^-)$ denotes the spin averaged mass of the $\rho$-mode excited states of $\Lambda_c$.
We assign the $\Lambda_c$ at $M(\Lambda_c,3/2^-)=2939.60$ MeV\cite{PDG} as a $\rho$-mode excited state, and 
assume that the splitting between $1/2^-$ and $3/2^-$ is the same as the lower ($\lambda$-mode) $\Lambda_c$ baryons, {\it i.e.}, $M(\Lambda_c;3/2^-)-M(\Lambda_c,1/2^-)=35.86$ MeV.
The other diquark masses are given by the mass formulas (\ref{mass1})--(\ref{mass5}) with $\epsilon=2/3$.

\begin{table}[tb]
  \centering
    \caption{Diquark masses and parameters of the chiral effective Lagrangian determined with each potential model. The values for scalar and pseudoscalar diquarks are taken from Ref.~\cite{scalar}. } 
  \begin{tabular}{ l  | c | c | c  } \hline\hline 
        \multicolumn{1}{ l | }{} & \multicolumn{3}{ c }{Potential model} \\ \cline{2 - 4} 
      & ~~Y-pot.~~\cite{yoshida} & ~~S-pot.~~\cite{silvestre} & ~~B-pot.~~\cite{barnes}   \\ \hline \hline
\multicolumn{4}{ l }{ Masses of S, P diquarks ($\rm MeV$) \cite{scalar}} \\ \hline
    $ M_{qq} (0^+)~ $ & 725  &725 &725   \\
    $ M_{qs} (0^+)~ $ & 942 &977 & 983   \\
    $ M_{qq} (0^-)~ $ & 1406 & 1484 &1496    \\ 
    $ M_{qs} (0^-)~ $ & 1271 & 1331 & 1341    \\  \hline   
\multicolumn{4}{ l }{ Masses of V, A diquarks ($\rm MeV$)} \\ \hline
    $ M_{qq} (1^+)~ $ & 973  &1013 &1019   \\
    $ M_{qs} (1^+)~ $ & 1116 &1170 & 1179   \\
    $ M_{ss} (1^+)~ $ & 1242 &1309 & 1320   \\ 
    $ M_{qq} (1^-)~ $ & 1447 & 1527 &1540    \\ 
    $ M_{qs} (1^-)~ $ & 1776 & 1883 & 1901    \\  \hline   
\multicolumn{4}{ l }{Parameters in $\mathcal{L}_S$ ({$\rm MeV^2$}) \cite{scalar}} \\ \hline    
    $~~m_{S0}^2~$ &$ (1119)^2$ &$ (1168)^2$&$ (1176)^2$  \\
    $~~m_{S1}^2~$ &$ (690)^2$ &$ (746)^2$&$ (754)^2$  \\
    $~~m_{S2}^2~$ &$-(258)^2$ &$-(298)^2$&$-(303)^2$  \\ \hline
\multicolumn{4}{ l }{Parameters in $\mathcal{L}_V$ ({$\rm MeV^2$})} \\ \hline    
    $~~m_{V0}^2~$ &$ (708)^2$ &$ (714)^2$&$ (714)^2$  \\
    $~~m_{V1}^2~$ &$-(757)^2$ &$-(808)^2$&$-(816)^2$  \\
    $~~m_{V2}^2~$ &$ (714)^2$ &$ (765)^2$&$ (773)^2$  \\ \hline \hline        
  \end{tabular}
  \label{diquarkmass}
\end{table}

The masses of the vector and axial-vector diquarks in the Y-potential model are smaller than those with the other two potentials. This behavior comes from the difference of the Coulomb interaction. Since the reduced mass of diquark and charm quark is about 600 MeV, the strength of the Coulomb potential, $\alpha$, in the Y-potential model is about 0.1 and is much smaller than that in the other models. The small value of $\alpha$ makes the binding energy smaller and the required diquark masses become smaller.
Note that for each potential model, the diquark masses satisfy the inequality Eq.~(\ref{in}).

\subsection{Strengths of the spin-dependent potentials}

The coefficients of the spin-spin potential, $\kappa_Q$, and the spin-orbit and tensor forces, $\eta$, are determined by known masses of the low-lying heavy baryons and summarized in Table~\ref{paraall}. 
We take the strengths of the spin-spin potential, $\kappa_c$ and $\kappa_b$, independently.
They are determined so as to reproduce the mass splittings, 
$M(\Sigma_c,1/2^+)=2453.54$ MeV and $M(\Sigma_c,3/2^+)=2518.13$ MeV, and
$M(\Sigma_b,1/2^+)=5813.10$ MeV and $M(\Sigma_b,3/2^+)=5832.53$ MeV, respectively.

The parameter $\eta$ for $V_{so}$ and $V_{ten}$ is determined from the splitting of the $\lambda$-mode
excitations of $\Lambda_c$.
Mass difference of $M(\Lambda_c,1/2^-)=2592.25$ MeV and $M(\Lambda_c,3/2^-)=2628.11$ MeV
comes from the spin-orbit interaction 
\begin{equation}
V_{so}(r) = \eta~ \frac{(1-e^{-\Lambda r})^2}{r^3} 
\left( \frac{2}{M_Q^2}+\frac{4}{M_d M_Q}\right) ~({\bm L}\cdot {\bm s}_Q),
\label{sopotlambda}
\end{equation}
which is reduced from Eqs.~(\ref{soplus}) and (\ref{sominus}) for the S diquark, $s_d=0$.
In order to reduce the number of the independent parameters, we assume that $\eta$ is independent
of the heavy-quark mass.
Furthermore, for all the three potentials, we use the same value $\eta=0.2494$ obtained with the Y-potential.

\subsection{Diquarks towards chiral restoration}

At high temperature and/or high baryon density, the QCD matter is expected to make a transition to a phase in which chiral symmetry is restored. In terms of the order parameter, the quark condensate $\langle \bar q q\rangle$ may vanish through the transition. It is of great interest to study how the diquarks are affected by the change of the chiral order parameter. For such a study, the effective theory based on the linear sigma model is most suitable though it may be too simplified. We here take the simplest approach for the chiral symmetry restoration by changing the chiral condensate $\langle \Sigma \rangle$ towards zero in due course.
\begin{figure}[b]
  \includegraphics[clip,width=1.0\columnwidth]{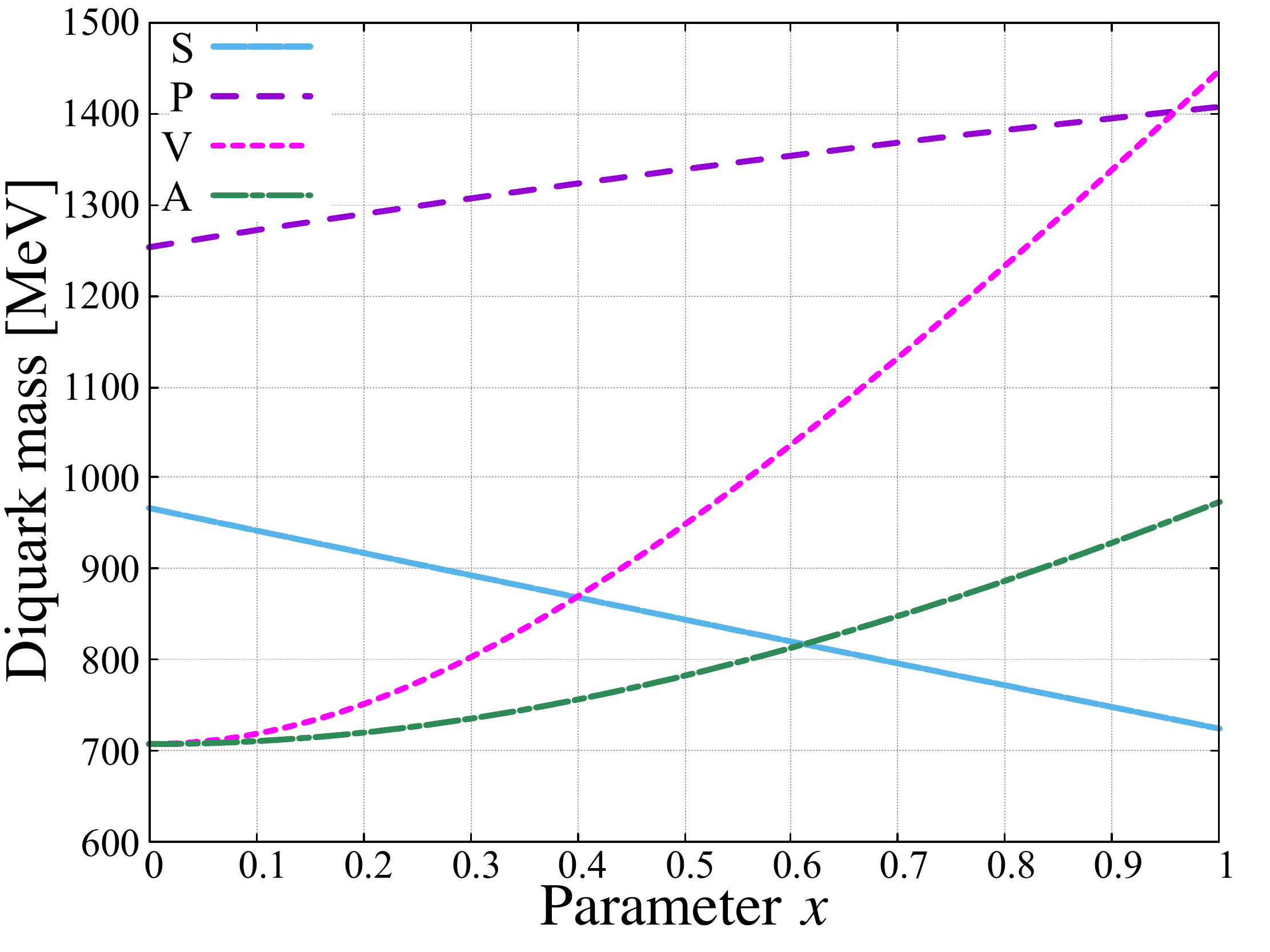}
  \caption{Dependence of the diquark masses on the chiral symmetry breaking parameter $x$.  }
  \label{chiralinvariant}
\end{figure}
In the present chiral effective theory, the diquark masses of the V and A diquarks will become degenerate to $m_{V0}$ when the chiral condensate $\langle \Sigma \rangle$ becomes zero. In Table \ref{diquarkmass}, one sees that $m_{V0}$ is about 700 MeV and is about 400 MeV lower than the mass of S and P diquarks in the chiral limit, $m_{S0}$. 
In contrast, the diquark masses in the ordinary vacuum is aligned so that the A diquark is heavier than
the S diquark.
Let us vary $\langle \Sigma \rangle$ by multiplying a factor $x$ and change $x$ from the vacuum, $x=1$, to 
the chiral symmetry restored phase, $x=0$. Correspondingly, the mass formulas for the non-strange S and A diquarks 
are given by
\begin{eqnarray}
&& M(0^+) = \sqrt{m^2_{S0} - (x+\epsilon) m^2_{S1} - x^2 m^2_{S2}},\\
&& M(1^+) = \sqrt{m^2_{V0} + x^2(m^2_{V1} + 2m^2_{V2})}.
\end{eqnarray}

\begin{table*}[tb]
  \centering
    \caption{Masses of the ground-state baryons with the S and A diquark. The values with asterisk (*) denote the input values from PDG~\cite{PDG}.}
  \begin{tabular}{ l | c | c | c | c | c } \hline\hline
  \multicolumn{1}{ l |}{} & \multicolumn{1}{ l |}{} &
  \multicolumn{4}{ l }{{\bf Baryon masses in the ground state (MeV)}} \\ \cline{3 -6}    
        \multicolumn{1}{ l |}{Baryon ($J^P$)}
      & \multicolumn{1}{ c |}{Diquark ($J^P$)}
      & \multicolumn{3}{c |}{Potential model} 
      & \multicolumn{1}{c}{} \\ \cline{3 - 5} 
      &  &
       ~Y-pot.~\cite{yoshida} &  ~S-pot.~\cite{silvestre}  &  ~B-pot.~\cite{barnes}  &Experiment~\cite{PDG}  \\ \hline 
    $\Lambda_c$ ($1/2^+$)&S ($0^+$)   &2286* &2286* &2286* & 2286.46 \\
    $\Sigma_c$ ($1/2^+$) &A ($1^+$)   &2452 &2453 &2441 & 2453.54 \\
    $\Sigma_c$ ($3/2^+$)&A ($1^+$)   &2516 &2517 &2521 & 2518.13 \\
    $\Xi_c$   ($1/2^+$)      &S ($0^+$)   &2469* &2469* &2469* & 2469.42 \\
    $\Xi_c' $  ($1/2^+$)     &A ($1^+$)   &2583 &2583 &2571 & 2578.80 \\
    $\Xi_c' $  ($3/2^+$)    &A ($1^+$)   &2642 &2643 &2647 & 2645.88 \\
    $\Omega_c$ ($1/2^+$)&A ($1^+$)   &2700 &2710 &2689 & 2695.20 \\
    $\Omega_c$ ($3/2^+$)&A ($1^+$)   &2755 &2758 &2762 & 2765.90 \\ 
    \hline
    $\Lambda_b$ ($1/2^+$)&S ($0^+$)   &5620* &5620* & $\ldots$ & 5619.60 \\
    $\Sigma_b$ ($1/2^+$) &A ($1^+$)   &5810 &5797 & $\ldots$ & 5813.10 \\
    $\Sigma_b$ ($3/2^+$) &A ($1^+$)   &5829 &5816 & $\ldots$ & 5832.53 \\
    $\Xi_b$   ($1/2^+$)      &S ($0^+$)   &5796 &5785 & $\ldots$ & 5794.45 \\
    $\Xi_b' $   ($1/2^+$)    &A ($1^+$)   &5934 &5914 & $\ldots$ & 5935.02 \\
    $\Xi_b' $   ($3/2^+$)    &A ($1^+$)   &5952 &5933 & $\ldots$ & 5953.82 \\
    $\Omega_b$ ($1/2^+$)&A ($1^+$)   &6047 &6022 & $\ldots$ & 6046.10 \\
    $\Omega_b$ ($3/2^+$)&A ($1^+$)   &6064 &6040 & $\ldots$ & $\ldots$ \\ \hline\hline
  \end{tabular}
  \label{tableg}
\end{table*}
\begin{table*}[tb]
  \centering
    \caption{Masses of the $\rho$-mode excited-state baryons with the P and V diquark. } 
  \begin{tabular}{ l | c | c | c | c | c } \hline\hline
  \multicolumn{1}{ l |}{} & \multicolumn{1}{ l |}{} &
  \multicolumn{4}{ l }{{\bf Baryon masses in the $\rho$-mode state (MeV)}} \\ \cline{3 -6}    
        \multicolumn{1}{ l |}{Baryon ($J^P$)}
      & \multicolumn{1}{ c |}{Diquark ($J^P$)}
      & \multicolumn{3}{c |}{Potential model} 
      & \multicolumn{1}{c}{} \\ \cline{3 - 5} 
     &  &
       ~Y-pot.~\cite{yoshida}  &  ~S-pot.~\cite{silvestre}  &  ~B-pot.~\cite{barnes}  &Experiment~\cite{PDG}  \\ \hline 
    $\Lambda_c$ ($1/2^-$)&P ($0^-$)   &2890 &2890 &2890 & $\ldots$ \\
    $\Lambda_c$ ($1/2^-$)&V ($1^-$)   &2894 &2892 &2881 & $\ldots$ \\
    $\Lambda_c$ ($3/2^-$)&V ($1^-$)   &2943 &2944 &2949 & (2939.60) \\
    $\Xi_c$ ($1/2^-$)&P ($0^-$)   &2765 &2758 &2758 & (2793.25) \\
    $\Xi_c$ ($1/2^-$)&V ($1^-$)   &3209 &3215 &3205 & $\ldots$ \\
    $\Xi_c$ ($3/2^-$)&V ($1^-$)   &3252 &3261 &3267 & $\ldots$ \\
    \hline
    $\Lambda_b$ ($1/2^-$)&P ($0^-$)   &6207 &6174 & $\ldots$ & $\ldots$ \\
    $\Lambda_b$ ($1/2^-$)&V ($1^-$)   &6233 &6197 & $\ldots$ & $\ldots$ \\
    $\Lambda_b$ ($3/2^-$)&V ($1^-$)   &6249 &6214 & $\ldots$ & $\ldots$ \\
    $\Xi_b$ ($1/2^-$)&P ($0^-$)   &6084 &6051 & $\ldots$ & $\ldots$ \\
    $\Xi_b$ ($1/2^-$)&V ($1^-$)   &6540 &6497 & $\ldots$ & $\ldots$ \\
    $\Xi_b$ ($3/2^-$)&V ($1^-$)   &6554 &6513 & $\ldots$ & $\ldots$ \\
\hline\hline
  \end{tabular}
  \label{tabler}
\end{table*}
\begin{table*}[htb]
  \centering
    \caption{Masses of the $\lambda$-mode excited-state baryons with the S and A diquark.} 
  \begin{tabular}{ l | c | c | c | c | c } \hline\hline
  \multicolumn{1}{ l |}{} & \multicolumn{1}{ l |}{} &
  \multicolumn{4}{ l }{{\bf Baryon masses in the $\lambda$-mode state (MeV)}} \\ \cline{3 -6}    
        \multicolumn{1}{ l |}{Baryon ($J^P$)}
      & \multicolumn{1}{ c |}{Diquark ($J^P$)}
      & \multicolumn{3}{c |}{Potential model} 
      & \multicolumn{1}{c}{} \\ \cline{3 - 5} 
     &  &
      ~Y-pot.~\cite{yoshida}  &  ~S-pot.~\cite{silvestre}  &  ~B-pot.~\cite{barnes} &Experiment~\cite{PDG}  \\ \hline 
    $\Lambda_c$ ($1/2^-$)&S ($0^+$)    &2589 &2676 & 2700 & (2592.25) \\
    $\Lambda_c$ ($3/2^-$)&S ($0^+$)    &2625 &2716 & 2749 & (2628.11) \\
    $\Sigma_c$ ($1/2^-$) &A ($1^+$)   &2717 &2807 &2830 &$\ldots$  \\
    $\Sigma_c$ ($1/2^-$) &A ($1^+$)   &2751 &2847 &2897 &$\ldots$  \\
    $\Sigma_c$ ($3/2^-$)&A ($1^+$)   &2781 &2881 &2919 &$\ldots$  \\
    $\Sigma_c$ ($3/2^-$)&A ($1^+$)   &2811 &2916 &2964 &$\ldots$  \\
    $\Sigma_c$ ($5/2^-$)&A ($1^+$)   &2844 &2953 &2994 &$\ldots$  \\
    $\Xi_c$ ($1/2^-$)&S ($0^+$)   &2754 &2852 & 2887 & (2793.25) \\
    $\Xi_c$ ($3/2^-$)&S ($0^+$)   &2787 &2890 & 2933 & (2818.50) \\
    $\Xi_c' $  ($1/2^-$)     &A ($1^+$)   &2845 &2942 &2971 &$\ldots$ \\
    $\Xi_c' $  ($1/2^-$)     &A ($1^+$)   &2876 &2979 &3033 &$\ldots$ \\
    $\Xi_c' $  ($3/2^-$)    &A  ($1^+$)  &2902 &3010 &3052 &$\ldots$ \\
    $\Xi_c' $  ($3/2^-$)    &A  ($1^+$)  &2926 &3037 &3090 &$\ldots$ \\
    $\Xi_c' $  ($5/2^-$)    &A  ($1^+$)  &2958 &3073 &3120 &$\ldots$ \\
    $\Omega_c$ ($1/2^-$)&A ($1^+$)   &2960 &3064 &3098 &$\ldots$ \\
    $\Omega_c$ ($1/2^-$)&A ($1^+$)   &2989 &3100 &3156 &$\ldots$ \\
    $\Omega_c$ ($3/2^-$)&A ($1^+$)   &3013 &3127 &3173 &$\ldots$ \\ 
    $\Omega_c$ ($3/2^-$)&A ($1^+$)   &3033 &3150 &3206 &$\ldots$ \\ 
    $\Omega_c$ ($5/2^-$)&A ($1^+$)   &3063 &3185 &3235 &$\ldots$ \\ 
    \hline
    $\Lambda_b$ ($1/2^-$)&S ($0^+$)   &5914 &6018 & $\ldots$ & (5912.20) \\
    $\Lambda_b$ ($3/2^-$)&S ($0^+$)   &5927 &6033 & $\ldots$ & (5919.92) \\
    $\Sigma_b$ ($1/2^-$) &A ($1^+$)   &6043 &6143 &$\ldots$ & $\ldots$ \\
    $\Sigma_b$ ($1/2^-$) &A ($1^+$)  &6065 &6171 &$\ldots$ & $\ldots$ \\
    $\Sigma_b$ ($3/2^-$)&A ($1^+$)   &6079 &6187 &$\ldots$ & $\ldots$ \\
    $\Sigma_b$ ($3/2^-$)&A ($1^+$)   &6117 &6234 &$\ldots$ & $\ldots$ \\
    $\Sigma_b$ ($5/2^-$)&A ($1^+$)   &6129 &6247 &$\ldots$ & $\ldots$ \\
    $\Xi_b$ ($1/2^-$)&S ($0^+$)   &6069 &6178 & $\ldots$ & $\ldots$ \\
    $\Xi_b$ ($3/2^-$)&S ($0^+$)   &6080 &6192 & $\ldots$ & $\ldots$ \\
    $\Xi_b' $  ($1/2^-$)     &A ($1^+$)   &6164 &6269 &$\ldots$ & $\ldots$\\
    $\Xi_b' $  ($1/2^-$)     &A ($1^+$)   &6183 &6293 &$\ldots$ &$\ldots$ \\
    $\Xi_b' $  ($3/2^-$)    &A ($1^+$)   &6195 &6307 &$\ldots$ & $\ldots$\\
    $\Xi_b' $  ($3/2^-$)    &A ($1^+$)   &6227 &6346 &$\ldots$ & $\ldots$\\
    $\Xi_b' $  ($5/2^-$)    &A ($1^+$)   &6238 &6359 &$\ldots$ & $\ldots$\\
    $\Omega_b$ ($1/2^-$)&A ($1^+$)   &6273 &6383 &$\ldots$ & $\ldots$\\
    $\Omega_b$ ($1/2^-$)&A ($1^+$)   &6290 &6404 &$\ldots$ & $\ldots$\\
    $\Omega_b$ ($3/2^-$)&A ($1^+$)   &6301 &6418 &$\ldots$ & $\ldots$\\ 
    $\Omega_b$ ($3/2^-$)&A ($1^+$)   &6329 &6452 &$\ldots$ & $\ldots$\\ 
    $\Omega_b$ ($5/2^-$)&A ($1^+$)   &6339 &6464 &$\ldots$ & $\ldots$\\ 
\hline\hline
  \end{tabular}
  \label{tablel}
\end{table*}

From the numerical values in all the potential models in Table \ref{diquarkmass}, 
we find that the $m_{S1}$ term is dominant
in the S diquark mass, while the $m_{V2}$ term is dominant in the A diquark.
Fig.~\ref{chiralinvariant} illustrates the masses of the $0^\pm$ and $1^\pm$ diquarks as functions of $x$ 
for the Y-potential.
One clearly sees that the mass of the S diquark will increase when we change $x$ from 1 towards 0,
for the Y-potential, starting from 725 MeV at $x=1$ reaching 967 MeV at $x=0$.
On the other hand, the mass of the A diquark will decrease from 973 MeV ($x=1$) to 708 MeV ($x=0$).
Thus there is a crossing in between. Namely, the inversion of the S and A 
diquarks will occur when the chiral symmetry is restored at finite temperature/baryon density.
In our Y-potential parameter set, the crossing occurs at around $x=0.6$.

It is also noted that at the chiral restoration, the masses of the non-strange S and P diquarks
are not degenerate due to the contribution of the strange quark mass induced by the $U_A(1)$ anomaly.
Here, for simplicity, we assume that the $U_A(1)$ anomaly stays intact under the chiral symmetry restoration.

This behavior of the diquark masses towards restoration of chiral symmetry may change
the baryon spectrum significantly at finite temperature and/or baryon density. It may also
affect the appearance of color superconductivity at dense matter, where the A diquarks may be lighter than the S diquarks.

\section{Numerical results} \label{Sec_4}

\subsection{Spectrum of singly charmed baryons}

\begin{figure*}[tbh]
  \includegraphics[clip,width=2.0\columnwidth]{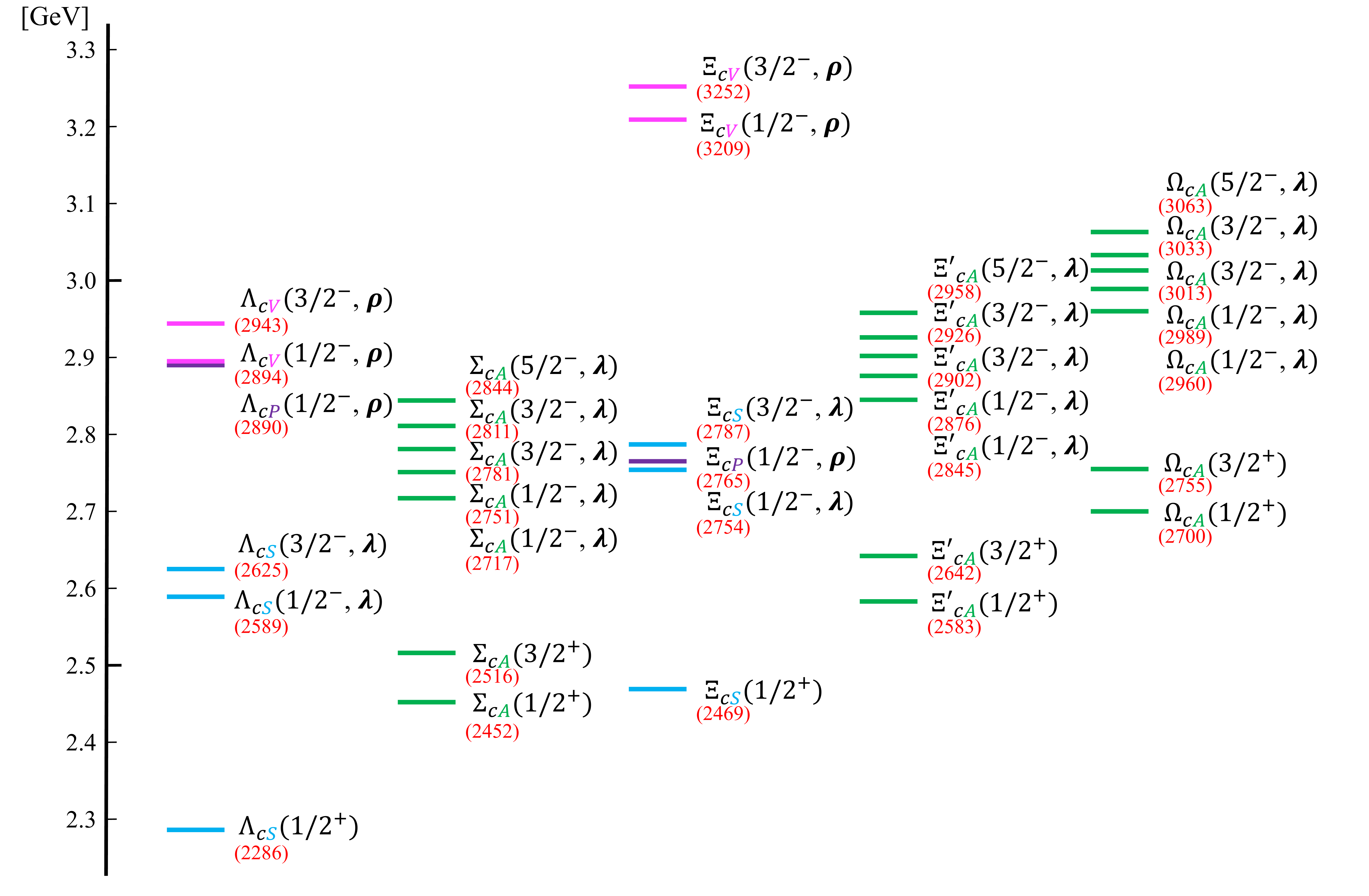}
  \caption{The energy spectra of singly charmed baryons from Tables \ref{tableg}, \ref{tabler}, and \ref{tablel}, given by the Y-potential. The colors of lines show the types of constituent diquarks as S (cyan), P (purple), V (magenta), and A (green). For the negative parity states, they are classified into the $\rho$-mode and the $\lambda$-mode with the symbols ${\bm \rho}$ and ${\bm \lambda}$, respectively.} 
  \label{figurecharm}
\end{figure*}

\begin{figure*}[tb]
  \includegraphics[clip,width=2.0\columnwidth]{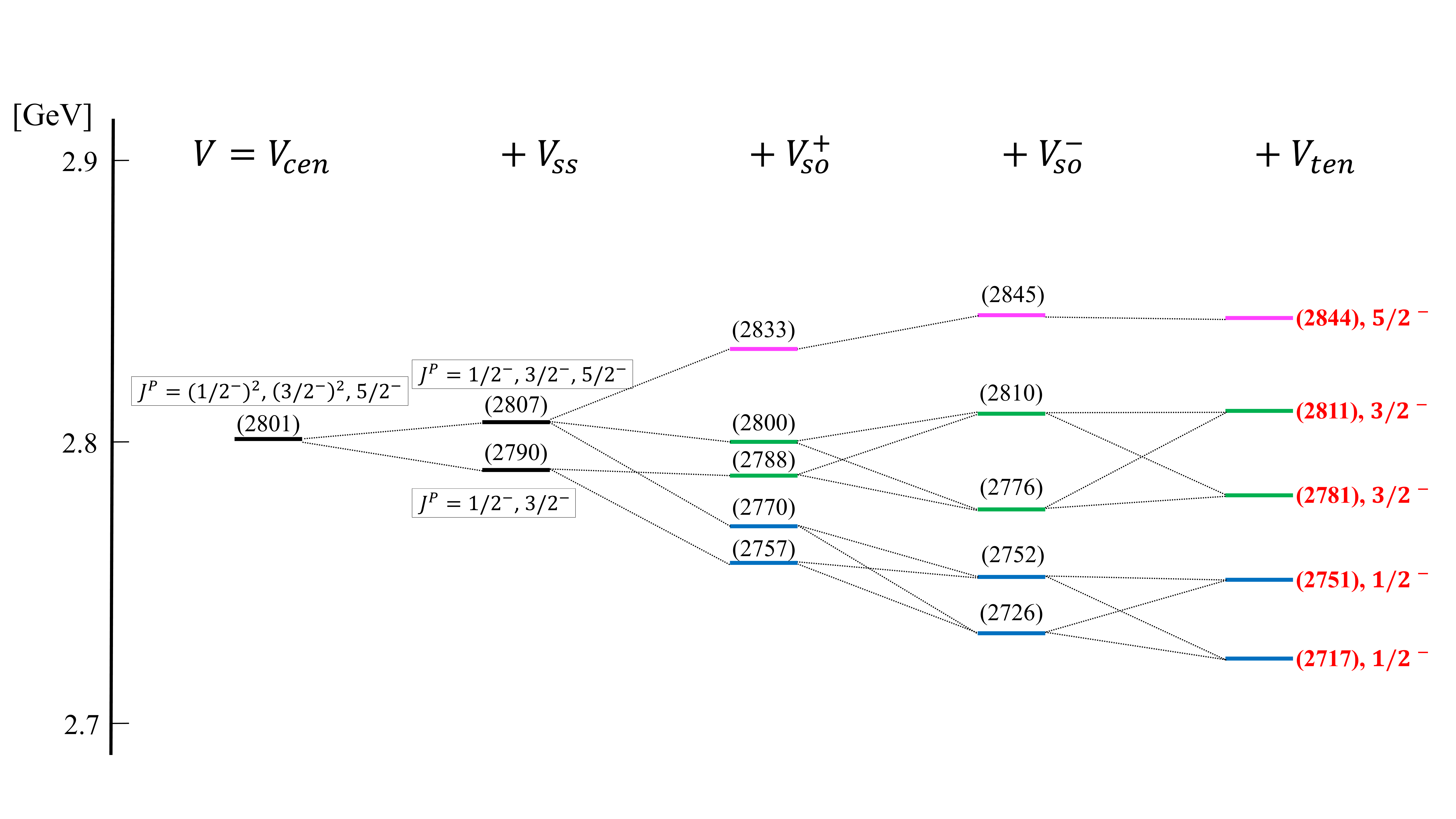}
  \caption{Contribution of the spin-dependent terms of the Y-potential to the energy splitting in the $\lambda$-mode states of $\Sigma_c$. The black dotted lines indicate the splitting and/or mixing of the states by adding the potential.} 
  \label{figurepotsplit}
\end{figure*}

\begin{figure*}[tbh]
  \includegraphics[clip,width=2.0\columnwidth]{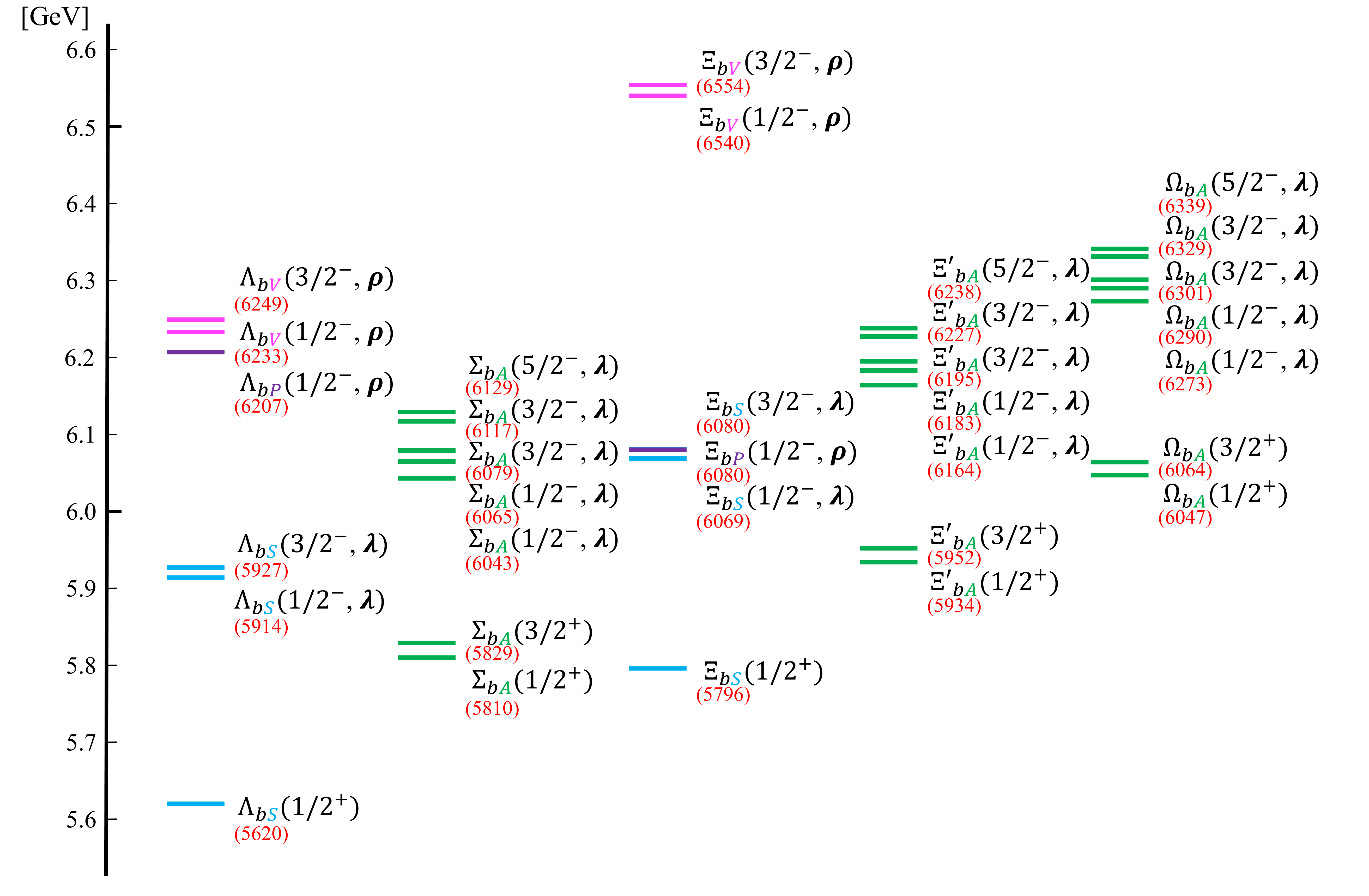}
  \caption{The energy spectra of singly bottom baryons from Tables \ref{tableg}, \ref{tabler}, and \ref{tablel}, given by the Y-potential. The colors of lines show the types of constituent diquarks as S (cyan), P (purple), V (magenta), and A (green). For the negative-parity states, they are classified into the $\rho$-mode and the $\lambda$-mode with the symbols ${\bm \rho}$ and ${\bm \lambda}$, respectively.} 
  \label{figurebottom}
\end{figure*}


Numerical results of masses of singly charmed baryons are summarized in the upper parts of Tables \ref{tableg}, \ref{tabler}, and \ref{tablel}, where we show the ground states, $\rho$-mode states, and $\lambda$-mode states, respectively.

First we explain which values are used to determine the diquark masses and the potential parameters. 
The values with asterisk in Table~\ref{tableg} are the experimental data taken from PDG \cite{PDG}.
Also, in Table~\ref{tabler}, the mass of $\Lambda_c(1/2^-)$ with the P diquark, 2890 MeV, is taken from Ref.~\cite{yoshida}. 
They are used to determine the parameter $C_c$ and the masses of the S and P diquarks \cite{scalar}.
In addition, for determining $\kappa_c$ and $\eta$, we used the observed mass differences between the $\Sigma_c(1/2^+)$ and $\Sigma_c(3/2^+)$ and between the $\lambda$-modes of the $\Lambda_c(1/2^-)$ and $\Lambda_c(3/2^-)$.

The energy spectra of singly charmed baryons from the Y-potential are shown in Fig. \ref{figurecharm}, where the difference in color indicates the type of the constituent diquarks inside baryons: S (cyan), P (purple), V (magenta), and A (green).
Note that the results of $\Lambda_c$ and $\Xi_c$ baryons including the S or P diquark were already reported in the previous work \cite{scalar}.
However, in Ref.~\cite{scalar}, the spin-orbit potential was neglected for simplicity, 
so that the $J^P=1/2^-$ and $3/2^-$ states in the $\lambda$-modes were degenerate. 

$\Lambda_c$ and $\Xi_c$ can include three types of constituent diquarks with flavor $\bar{\bm 3}$.
First we focus on the ground states and the $\lambda$-mode states with the S diquark and the $\rho$-mode states with the P diquark, which is explained in the previous work~\cite{scalar}.
In each state, the masses of singly charmed baryons including S diquark become heavier with increasing the number of constituent strange quarks. 
On the other hand, for the $\rho$-modes of baryons with the P diquark, the non-strange baryon is heavier than that of  including one strange quark as $M_\rho(\Lambda_c, 1/2^-) > M_\rho(\Xi_c, 1/2^-)$. This is caused by the inverse-mass hierarchy of P diquarks, Eq.~(\ref{inverse}). 
Another diquark appearing in the spectrum of $\Lambda_c$ and $\Xi_c$ is the V diquark, which makes the heaviest $\rho$-mode states. 
The masses of $\Xi_c$ are about 300 MeV heavier than those of $\Lambda_c$. 
This coincides with the mass difference between the strange and non-strange V diquarks is $M_{qs}(1^-)-M_{qq}(1^-)=329$ MeV, as shown in Table \ref{diquarkmass}.

Next we discuss the spectra of $\Sigma_c$, $\Xi'_c$, and $\Omega_c$. 
These baryons include the A diquark, which gives both the ground states and $\lambda$-mode excited states.
In our model, the masses of these baryons become heavier as the total angular momentum $J$ is larger.
For instance, in the ground states, the masses for $J^P=3/2^+$ are about 50 MeV larger than those of $1/2^+$.
Also the masses of $\lambda$-modes with $J^P=1/2^-$, $3/2^-$, and $5/2^-$ are in order.

The mass differences between $\Sigma_c$ and $\Xi'_c$ with the same $J^P$ are about 130 MeV, and those between $\Xi'_c$ and $\Omega_c$ are about 120 MeV.
This behavior is regarded as a generalization of the conventional Gell-Mann--Okubo mass formula~\cite{gomass1, gomass2}.
Values of these mass differences are almost from those of A diquarks, $M_{qs}(1^+)-M_{qq}(1^+)=143$ MeV and $M_{ss}(1^+)-M_{qs}(1^+)= 126$ MeV, which are smaller than that of V diquarks ($M_{qs}(1^-)-M_{qq}(1^-)=329$ MeV). 
This is due to the negative value of the parameter $m_{V1}^2 < 0$, where the square mass difference between the A diquarks in Eq.~(\ref{axial-sqmassdiff}) becomes smaller than that between the V diquarks in Eq.~(\ref{vec-sqmassdiff}).

Comparing with each potential, there is no much difference in the ground state and the $\rho$-modes according to Tables \ref{tableg} and \ref{tabler}. However, in the $\lambda$-mode state in Table \ref{tablel}, the masses given by the S- and B-potentials are approximately 100 MeV heavier than those of the Y-potential.
In other words, the excitation energy from the ground state to the $\lambda$-mode is approximately 200--300 MeV for the Y-potential and 300--400 MeV for the S- and B-potentials.
This difference is caused by the coefficients of the Coulomb potential, $\alpha$, whose value is approximately $0.1$ for the Y-potential and is smaller than the other two potentials.
As a result, the wave function obtained from the Y-potential becomes broader, and its difference between the ground state and $\lambda$-mode becomes smaller.
Therefore, the excitation energy decreases as $\alpha$ becomes smaller, as explained in the previous work~\cite{scalar}. 
According to the Table \ref{tablel}, about the masses of the $\lambda$-mode of $\Lambda_c(1/2^-)$ and $\Lambda_c(3/2^-)$, the numerical results by the Y-potential are mostly closest to the experimental data from PDG \cite{PDG} than the other two potentials. From now, we mainly use the Y-potential for the two-body calculation.

Lastly, we discuss the splitting of energy spectrum by the spin dependent potentials.
In Fig. \ref{figurepotsplit}, we sketch the splitting behaviors in the $\lambda$-modes of $\Sigma_c$ by adding the spin-spin ($V_{ss}$), spin-orbit ($V_{so}$), and tensor ($V_{ten}$) potentials. 
At first, two $J^P=1/2^-$, two $J^P=3/2^-$, and one $J^P=5/2^-$ states are degenerate, and the spin-spin potential splits these five degenerate states into two parts. 
The lower states with $J^P=1/2^-$ and $3/2^-$ are degenerate, and the higher states with $J^P=1/2^-, 3/2^-$, and $5/2^-$ also do.
Here, the mass difference between the lower and higher parts is about 20 MeV.
After this, the symmetric term of spin-orbit potential $V_{so}^+$ splits these five states completely, and the antisymmetric term $V_{so}^-$ mixes the states with the same $J^P$. We can clearly see that the splitting by the spin-orbit potential changes the ordering of total angular momentum $J$ into $1/2^- < 1/2^- < 3/2^- < 3/2^- < 5/2^-$, and the splitting range is approximately 120 MeV. Finally, the contribution from the tensor potential is very small.

\subsection{Spectrum of singly bottom baryons}

Similarly to singly charmed baryons, the numerical results for singly bottom baryons are summarized in the lower parts of Tables \ref{tableg}, \ref{tabler}, and \ref{tablel}.
Here only the mass of $\Lambda_b (1/2^+)$ including the S diquark is the input value for determining the parameter $C_b$ \cite{scalar}.
The ground states of $\Sigma_b (1/2^+)$ and $\Sigma_b (3/2^+)$ are used to determine the parameter $\kappa_b$.

The energy spectra of singly bottom baryons from the Y-potential are shown in Fig. \ref{figurebottom}.
We can see that the spectrum of singly bottom baryons are similar to those of singly charmed baryon.
In particular, we can find the inverse-mass hierarchy of the $\rho$-mode excitations as $M_{\rho}(\Lambda_b, 1/2^-) > M_{\rho}(\Xi_b, 1/2^-)$, which is caused by the inverse hierarchy of the pseudoscalar diquarks in Eq.~(\ref{inverse}).

By comparing Figs. \ref{figurecharm} and \ref{figurebottom}, one sees the heavy-quark mass dependence of masses of singly heavy baryons.
The spin-spin potential $V_{ss}$, Eqs.~(\ref{ssb}) and (\ref{ssys}), and the tensor potential $V_{ten}$, Eq.~(\ref{tenpot}), are inversely proportional to both the diquark mass and heavy-quark mass ($V_{ss,ten} \propto 1/M_d M_Q$).
Also, as in Eqs. (\ref{soplus}) and (\ref{sominus}), some terms of the spin-orbit potential $V_{so}$ are inversely proportional to the heavy-quark mass $M_Q$.
In the heavy-quark limit ($M_Q \rightarrow \infty$), these terms become zero, and the surviving term in the spin-orbit potential is  
\begin{eqnarray}
V_{so}(r) = \eta \frac{(1-e^{\Lambda r})^2}{r^3} \frac{2}{M_d^2} ({\bm L} \cdot {\bm s}_d),
\end{eqnarray}
which is independent of the heavy-quark spin ${\bm s}_Q$.
This limit induces characteristic structures of hadron mass spectra: (i) the heavy-quark spin (HQS) doublet with $J=j \pm 1/2$, which is a pair of two states characterized by the angular momentum $j$ of the light components and the heavy-quark spin $1/2$, and (ii) the HQS singlet, where the angular momentum of light components is zero ($j=0$). 

For example, the ground states of $\Sigma_Q$, $\Xi'_Q$, and $\Omega_Q$ with the A diquark consist of two states with $J=1/2$ and $3/2$ using the diquark spin $j=1$.
The mass difference between $J^P=1/2^+$ and $3/2^+$ is approximately 60 MeV for the charmed baryons, while that for the bottom baryons is about 20 MeV.
Such a smaller mass difference for the bottom baryons is an example of the HQS doublet structure.
In our spectra, similar HQS doublet structures can be also found in the $\lambda$ modes of $\Lambda_Q$ and $\Xi_Q$ with the S diquark (using the orbital angular momentum $j=1$) and the $\rho$ modes of $\Lambda_Q$ and $\Xi_Q$ with the V diquark (using the diquark spin $j=1$).
Furthermore, the spectra of the $\lambda$ modes of $\Sigma_Q$, $\Xi'_Q$, and $\Omega_Q$ consist of two HQS doublets ($J=1/2$ and $3/2$ using $j=1$, and $J=3/2$ and $5/2$ using $j=2$) and one HQS singlet ($J=1/2$ using $j=0$ due to the cancellation of the diquark spin and orbital angular momentum).

\subsection{Spectrum of positive-parity excited states}

\begin{table*}[tb]
  \centering
    \caption{Masses of radially-excited-state baryons with the number of node $n=1$ and the $D$-wave excited-state baryons with the orbital angular momentum of the $\lambda$-coordinate $L=2$. Y-potential is used for this calculation. }
  \begin{tabular}{ c  | r | r | c | c ||  c  | r | r | c | c } \hline\hline 
     Charmed baryon ($J^P$) & \quad$n$ & \quad$L$ & Mass (MeV) & Experiment~\cite{PDG}
       &Bottom Baryon ($J^P$) & \quad$n$ & \quad$L$ & Mass (MeV) & Experiment~\cite{PDG}\\ \hline
  $\Lambda_c (1/2^+)$ & 1 &$0$&  2825 & (2765) &   $\Lambda_b (1/2^+)$ & 1 &$0$&  6121 & \ldots \\ 
  $\Lambda_c (3/2^+)$ & 0 &$2$&  2869 & 2856.10 &   $\Lambda_b (3/2^+)$ & 0 &$2$&  6172 & 6146.20 \\ 
  $\Lambda_c (5/2^+)$ & 0 &$2$&  2897 & 2881.63 &  $\Lambda_b (5/2^+)$ & 0 &$2$&  6178 & 6152.50 \\ \hline
  $\Sigma_c (1/2^+)$ & 1 &$0$&  2974 & \ldots &   $\Sigma_b (1/2^+)$ & 1 &$0$&  6274 & \ldots \\ 
  $\Sigma_c (1/2^+)$ & 0 &$2$&  3016 & \ldots &   $\Sigma_b (1/2^+)$ & 0 &$2$&  6304 & \ldots\\ 
  $\Sigma_c (3/2^+)$ & 1 &$0$&  3013 & \ldots &   $\Sigma_b (3/2^+)$ & 1 &$0$&  6286 & \ldots\\ 
  $\Sigma_c (3/2^+)$ & 0 &$2$&  3043 & \ldots &   $\Sigma_b (3/2^+)$ & 0 &$2$&  6316 & \ldots\\ 
  $\Sigma_c (3/2^+)$ & 0 &$2$&  3053 & \ldots &   $\Sigma_b (3/2^+)$ & 0 &$2$&  6330 & \ldots\\ 
  $\Sigma_c (5/2^+)$ & 0 &$2$&  3076 & \ldots &   $\Sigma_b (5/2^+)$ & 0 &$2$&  6341 & \ldots\\ 
  $\Sigma_c (5/2^+)$ & 0 &$2$&  3094 & \ldots &   $\Sigma_b (5/2^+)$ & 0 &$2$&  6365 & \ldots\\ 
  $\Sigma_c (7/2^+)$ & 0 &$2$&  3115 & \ldots &   $\Sigma_b (7/2^+)$ & 0 &$2$&  6373 & \ldots\\  \hline
  $\Xi_c (1/2^+)$ & 1 &$0$&  2976 & \ldots &   $\Xi_b (1/2^+)$ & 1 &$0$&  6260 & \ldots\\ 
  $\Xi_c (3/2^+)$ & 0 &$2$&  3017 & \ldots &   $\Xi_b (3/2^+)$ & 0 &$2$&  6307 & \ldots\\ 
  $\Xi_c (5/2^+)$ & 0 &$2$&  3043 & \ldots &   $\Xi_b (5/2^+)$ & 0 &$2$&  6313 & \ldots\\ \hline
  $\Xi'_c (1/2^+)$ & 1 &$0$&  3088 & \ldots &   $\Xi'_b (1/2^+)$ & 1 &$0$&  6381 & \ldots\\ 
  $\Xi'_c (1/2^+)$ & 0 &$2$&  3131 & \ldots &   $\Xi'_b (1/2^+)$ & 0 &$2$&  6411 & \ldots\\ 
  $\Xi'_c (3/2^+)$ & 1 &$0$&  3124 & \ldots &   $\Xi'_b (3/2^+)$ & 1 &$0$&  6392 & \ldots\\ 
  $\Xi'_c (3/2^+)$ & 0 &$2$&  3155 & \ldots &   $\Xi'_b (3/2^+)$ & 0 &$2$&  6423 & \ldots\\ 
  $\Xi'_c (3/2^+)$ & 0 &$2$&  3164 & \ldots &   $\Xi'_b (3/2^+)$ & 0 &$2$&  6434 & \ldots\\ 
  $\Xi'_c (5/2^+)$ & 0 &$2$&  3186 & \ldots &   $\Xi'_b (5/2^+)$ & 0 &$2$&  6444 & \ldots\\ 
  $\Xi'_c (5/2^+)$ & 0 &$2$&  3200 & \ldots &   $\Xi'_b (5/2^+)$ & 0 &$2$&  6465 & \ldots\\ 
  $\Xi'_c (7/2^+)$ & 0 &$2$&  3221 & \ldots &   $\Xi'_b (7/2^+)$ & 0 &$2$&  6472 & \ldots\\  \hline
  $\Omega_c (1/2^+)$ & 1 &$0$&  3194 & \ldots &   $\Omega_b (1/2^+)$ & 1 &$0$&  6480 & \ldots\\ 
  $\Omega_c (1/2^+)$ & 0 &$2$&  3236 & \ldots &   $\Omega_b (1/2^+)$ & 0 &$2$&  6511 & \ldots\\ 
  $\Omega_c (3/2^+)$ & 1 &$0$&  3227 & \ldots &   $\Omega_b (3/2^+)$ & 1 &$0$&  6491 & \ldots\\ 
  $\Omega_c (3/2^+)$ & 0 &$2$&  3259 & \ldots &   $\Omega_b (3/2^+)$ & 0 &$2$&  6522 & \ldots\\ 
  $\Omega_c (3/2^+)$ & 0 &$2$&  3268 & \ldots &   $\Omega_b (3/2^+)$ & 0 &$2$&  6532 & \ldots\\ 
  $\Omega_c (5/2^+)$ & 0 &$2$&  3288 & \ldots &   $\Omega_b (5/2^+)$ & 0 &$2$&  6541 & \ldots\\ 
  $\Omega_c (5/2^+)$ & 0 &$2$&  3300 & \ldots &   $\Omega_b (5/2^+)$ & 0 &$2$&  6559 & \ldots\\ 
  $\Omega_c (7/2^+)$ & 0 &$2$&  3320 & \ldots &   $\Omega_b (7/2^+)$ & 0 &$2$&  6567 & \ldots\\  \hline\hline
  \end{tabular}
  \label{tablepositive}
\end{table*}

\begin{figure*}[htb]
  \includegraphics[clip,width=2.0\columnwidth]{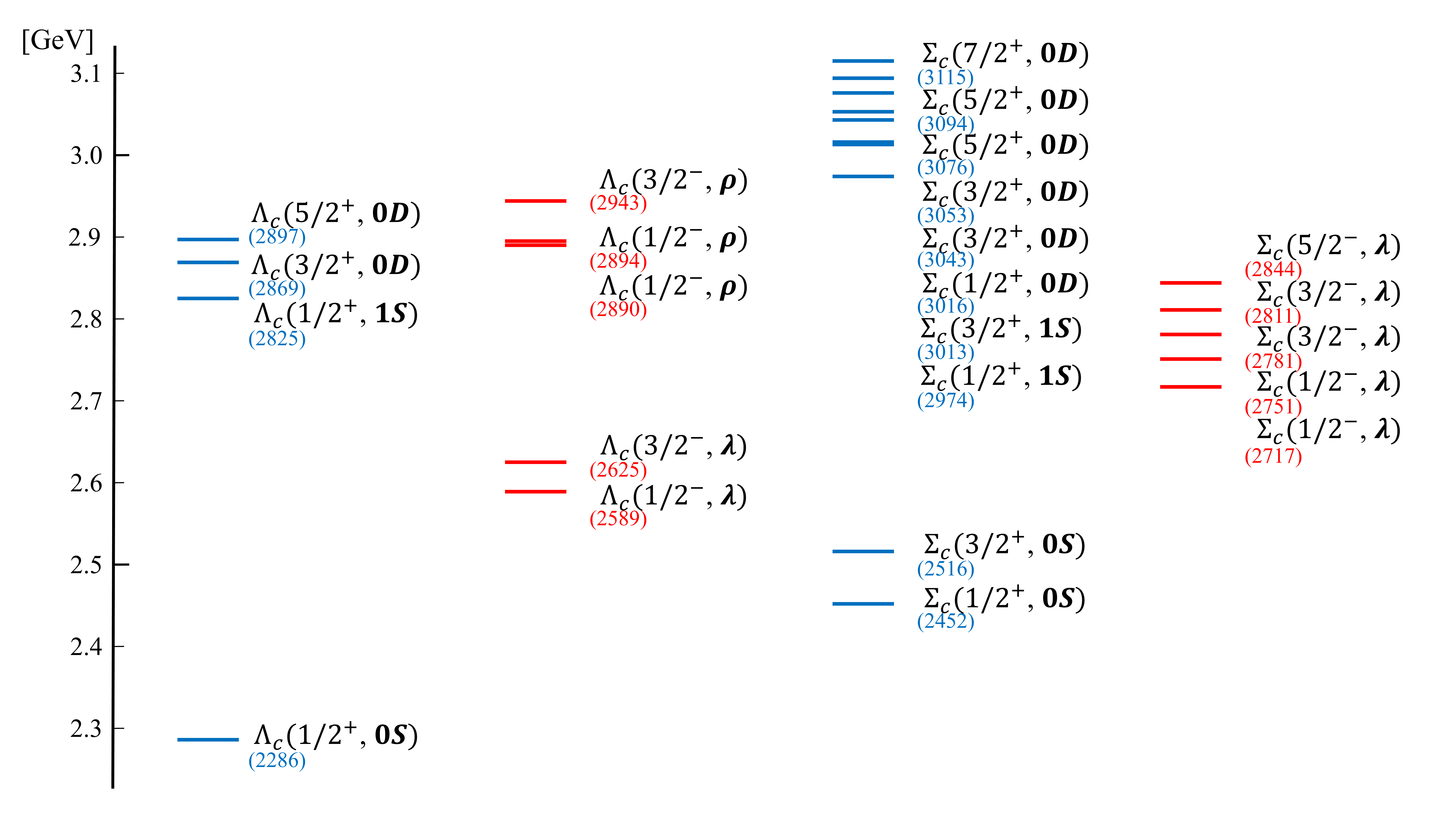}
  \caption{The energy spectra of $\Lambda_c$ and $\Sigma_c$ from Tables \ref{tableg}, \ref{tabler}, \ref{tablel}, and \ref{tablepositive}, given by the Y-potential. The colors of lines show the parities (Blue: positive parity. Red: negative parity.). For the blue spectra, they are classified by $n=0$ or $1$ and $L=0(S)$ or $2(D)$, expressed as the symbol $nL=0S, 1S$, and $0D$.}
\label{figurecharm2}
\end{figure*}
\begin{figure*}[htb]
  \includegraphics[clip,width=2.0\columnwidth]{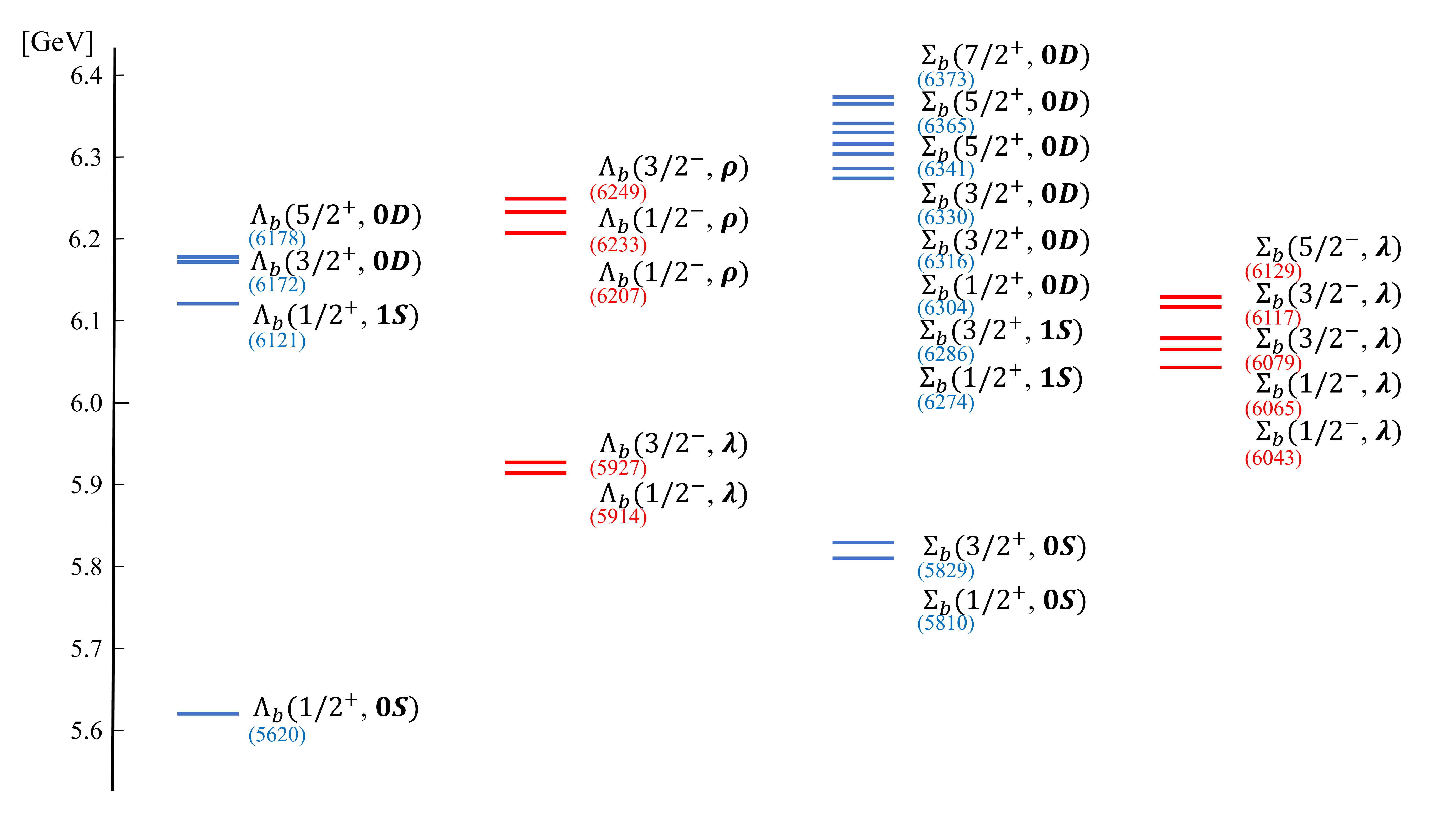}
  \caption{The energy spectra of $\Lambda_b$ and $\Sigma_b$ from Tables \ref{tableg}, \ref{tabler}, \ref{tablel}, and \ref{tablepositive}, given by the Y-potential.
  }
\label{figurebottom2}
\end{figure*}

In the previous subsections, we have discussed the (positive-parity) ground states and negative-parity excited states.
Here, we study two types of positive-parity excited states.
The first is the radially excited states, where the number of nodes is $n=1$ (i.e., the principal quantum number is $2$), and the orbital angular momentum is zero ($L=0$).
The second is the $D$-wave excited states, where the orbital angular momentum in the $\lambda$ coordinate is $L=2$.
These states have positive parity and are composed of the S or A diquarks.
The numerical results from the Y-potential are summarized in Table~\ref{tablepositive}, where the left and right parts show the singly charmed and bottom baryons, respectively.
Among them, we show the energy spectra of $\Lambda_{c/b}$ and $\Sigma_{c/b}$ in Figs.~\ref{figurecharm2} and \ref{figurebottom2}, respectively.

From these figures,  we find that the masses of the radially excited states are lower than those of the corresponding $D$-wave excited states. For example, $\Lambda_c(1/2^+,1S)$ is lighter than $\Lambda_c(3/2^+,0D)$ and $\Lambda_c(5/2^+,0D)$.
However, the mass difference between these states are not so much large.
Also, for the $1S$ and $0D$ states of $\Lambda_{c/b}$ and $\Sigma_{c/b}$, as the total angular momentum becomes larger, the mass becomes heavier.
For the singly bottom baryons, we can clearly see the HQS doublet structures of the positive-parity excited states.
As a result of this structure, the range of the energy spectrum of bottom baryons tends to shrink.

We compare the masses of the positive-parity excited states and those of negative-parity $\rho$- and $\lambda$-modes.
For $\Lambda_{c/b}$, we can see that their masses are close to each other, and the $\rho$-modes are slightly heavier than the positive-parity excited states.
On the other hand, the $\lambda$-modes are lying in between the ground state and the positive-parity excited states.
This tendency for the $\lambda$-modes is also seen in the spectrum of $\Sigma_{c/b}$. 

\subsection{Comparison with experimental data}

\begin{figure*}[tp]
  \includegraphics[clip,width=2.0\columnwidth]{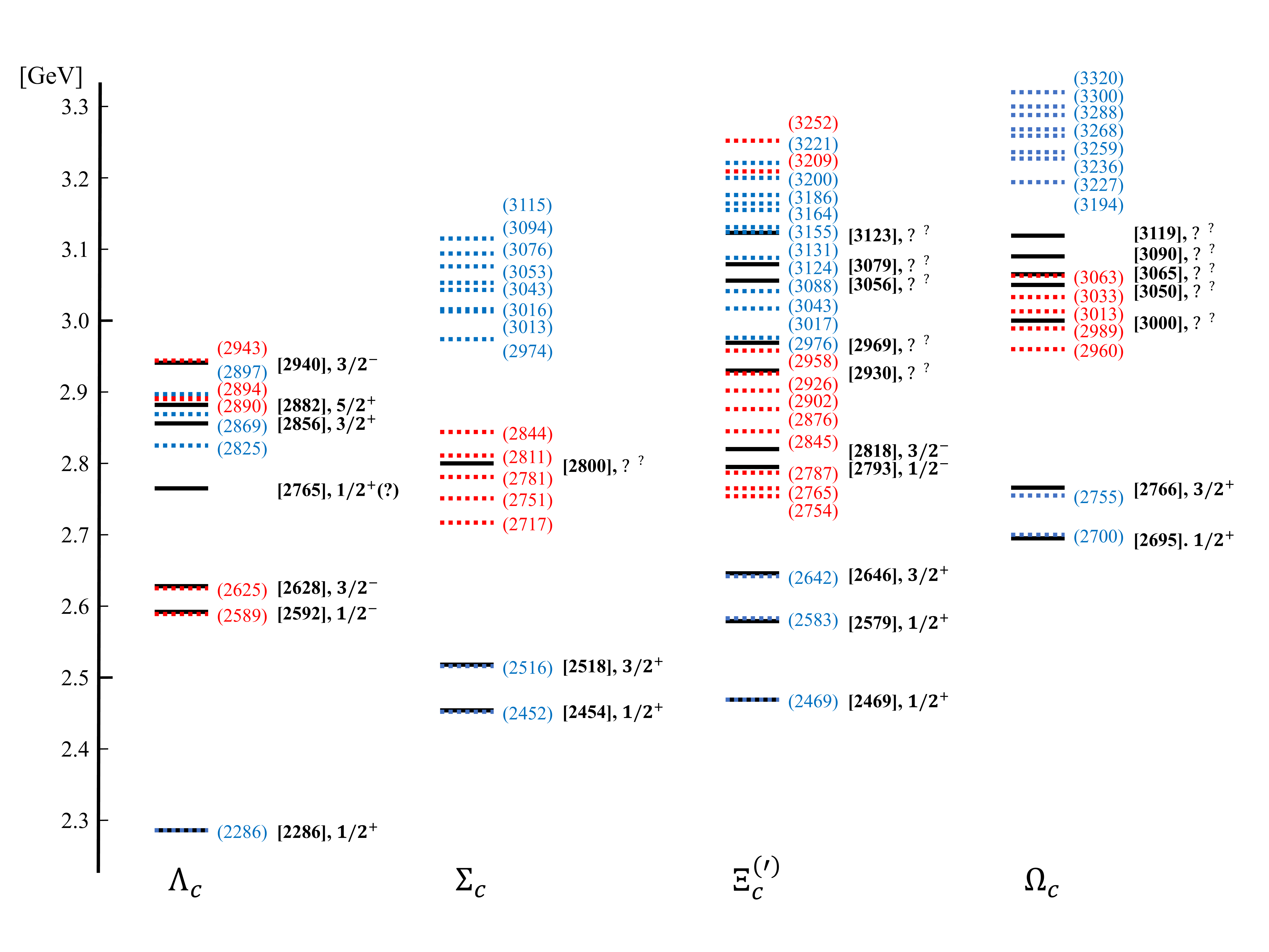}
  \caption{Energy spectra of singly charmed baryons from our model with the Y-potential (blue or red dotted lines), and from the experimental data in PDG~\cite{PDG} (black solid lines). Here we put the spectra of $\Xi_c$ and $\Xi'_c$ together.}
\label{figurecharmexp}
\end{figure*}

\begin{figure*}[tp]
  \includegraphics[clip,width=2.0\columnwidth]{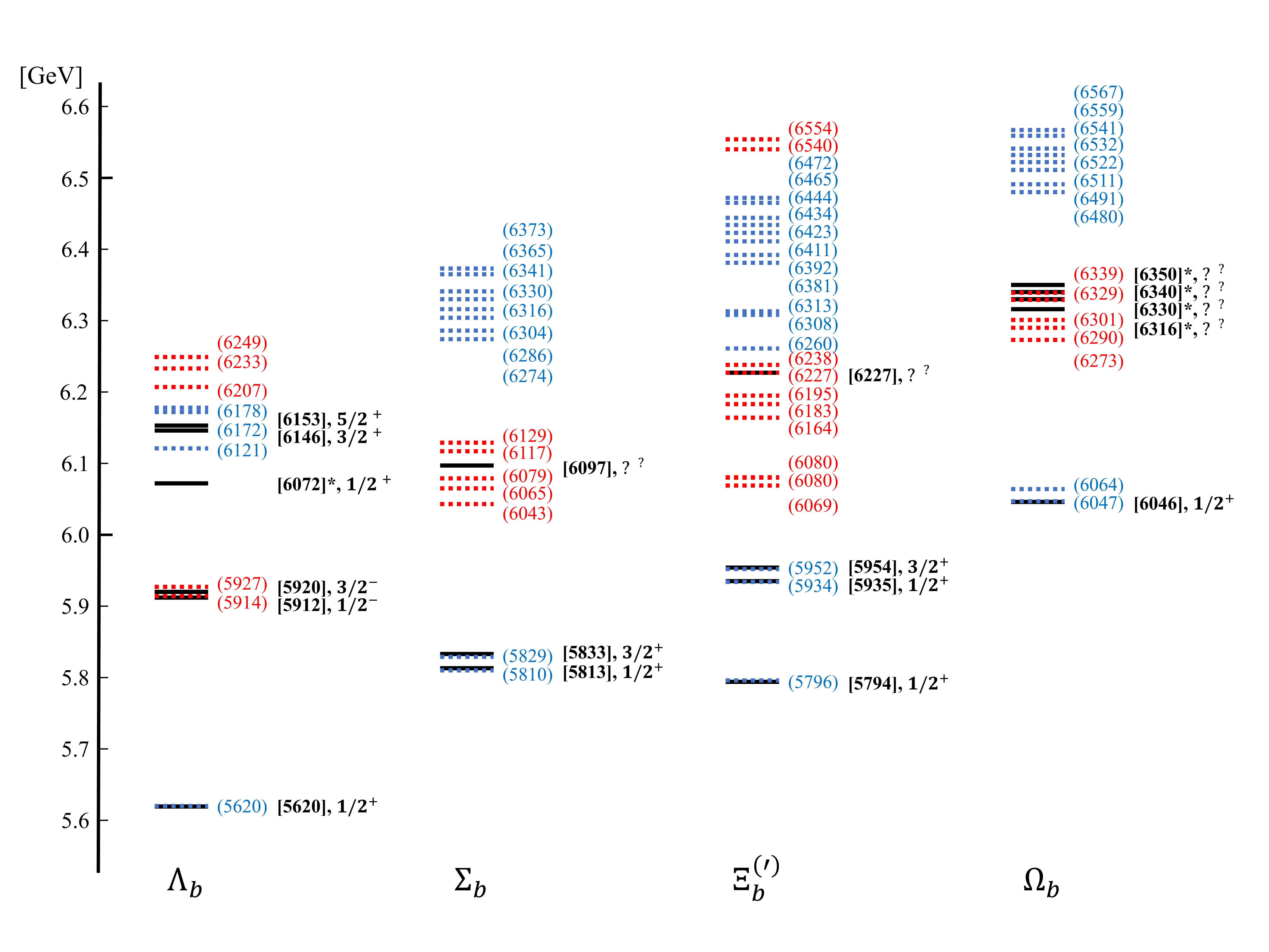}
  \caption{Energy spectra of singly bottom baryons from our model with the Y-potential (blue or red dotted lines), and from the experimental data in PDG~\cite{PDG} (black solid lines). Here we put the spectra of $\Xi_b$ and $\Xi'_b$ together. Note that $\Lambda_b (1/2^+)$ with the mass of 6072 MeV and $\Omega_b$ excited states with the asterisk (*) are the experimental values from Refs. \cite{lambdab6072, omegab6300}.}
\label{figurebottomexp}
\end{figure*}

Here, we compare the numerical results and experimental data.
In the last column of Tables \ref{tableg}, \ref{tabler}, \ref{tablel}, and \ref{tablepositive}, we show the masses of the baryons whose spin and parity are specified by PDG \cite{PDG}.
In Figs. \ref{figurecharmexp} and \ref{figurebottomexp}, we show the energy spectra of singly heavy baryons, where both our results and experimental data are shown.

The masses of the ground states of $\Lambda_{c/b} (1/2^+)$ and $\Xi_c (1/2^+)$ are the input values. Also the mass differences between the ground states of $\Sigma_{c/b} (1/2^+)$ and $\Sigma_{c/b} (3/2^+)$ and between the $\lambda$-mode states of $\Lambda_c (1/2^-)$ and $\Lambda_c (3/2^-)$ are the input values, where the obtained mass values are similar to the observed data.
Therefore the spectra of $\Xi^{(')}_b$ and $\Omega_{c/b}$ are the pure outputs from our model.

It is obvious that, in the ground states, all the mass values from our model are near the experimental values.
On the other hand, for the $\lambda$-mode states of $\Omega_{c/b}$, the masses from our model seem to be lower than the observed values.
This is because we have not adjusted the potential parameters of the Coulomb interaction term $\alpha$ and the linear confinement term $\lambda$ for our heavy-quark--diquark model. 

In Table \ref{tablepositive}, $\Lambda_{c/b} (3/2^+)$ and $\Lambda_{c/b} (5/2^+)$ are experimentally observed.
We expect that these states are the $D$-wave excited states ($L=2$) from the viewpoints of our diquark model.
Although the masses from our model are about 20 MeV larger than the experimental values, the mass splitting between the $3/2^+$ and $5/2^+$ states is almost same as the experimental value.

Next we comment on the $n=1$ excited states with $1/2^+$.
The mass of the first excited state of $\Lambda_c (1/2^+)$ is known to be 2765 MeV. Its isospin is determined by the Belle collaboration \cite{lambdac2765a}, while $J^P$ is not determined yet but may be assumed as $1/2^+$.
In addition, recently, the LHCb Collaboration have observed the first excited state of $\Lambda_b (1/2^+)$ whose mass is 6072 MeV \cite{lambdab6072}.
These states may be analogous to the Roper resonance in the nucleon spectrum.
Our results, $\Lambda_c(1/2^+)=2825$ MeV and $\Lambda_b(1/2^+)=6121$ MeV, are approximately 50 MeV heavier than the experimental masses. 
Thus, so far, we cannot provide an interpretation based on the heavy-quark--diquark picture.

For the other excited states $\Sigma_{c/b}$, $\Xi_{c/b}$, and $\Omega_{c/b}$, $J^P$ is still unknown, except for $\Xi_c (1/2^-)$ and $\Xi_c (3/2^-)$.
It may be difficult to determine the unknown total angular momenta from our model, but at least we can expect the signs of parity.
For instance, $\Sigma_c(2800)$, $\Sigma_b(6097)$, the five states of $\Omega_c$, and the four observed states of $\Omega_b$ may be $\lambda$-mode states with negative parity.
On the other hand, it is not conclusive to identify the quantum numbers of the five higher states of $\Xi_c^{(')}$ and $\Xi_b(6227)$.
Note that, recently, $\Xi_c^{(')}(2930)$ is reinterpreted as $\Xi_c(2923)$ and $\Xi_c(2939)$ \cite{Aaij:2020yyt}, and the spin and parity of $\Xi_c(2970)$ are determined as $1/2^+$ \cite{Moon:2020gsg}.
Also, the new state $\Xi_b(6100)$ is recently observed \cite{Sirunyan:2021vxz}.

\subsection{Comparison with three-quark model}

\begin{figure*}[tp]
  \includegraphics[clip,width=2.0\columnwidth]{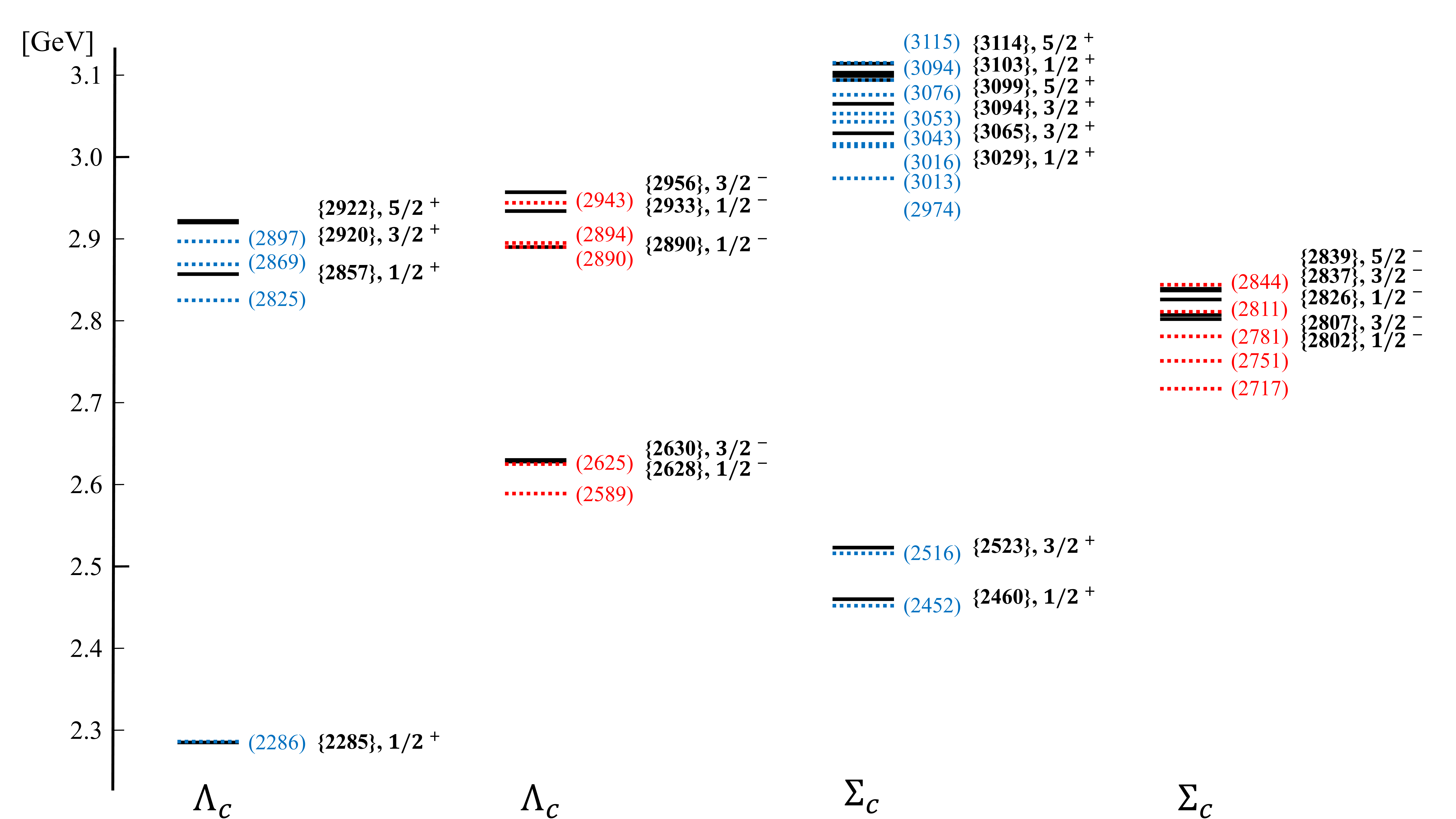}
  \caption{Energy spectra of $\Lambda_c$ and $\Sigma_c$ from our model with Y-potential (blue or red dotted lines), and from the three-quark model in Ref. \cite{yoshida} (black solid lines).}
\label{figurecharm3q}
\end{figure*}

\begin{figure*}[tp]
  \includegraphics[clip,width=2.0\columnwidth]{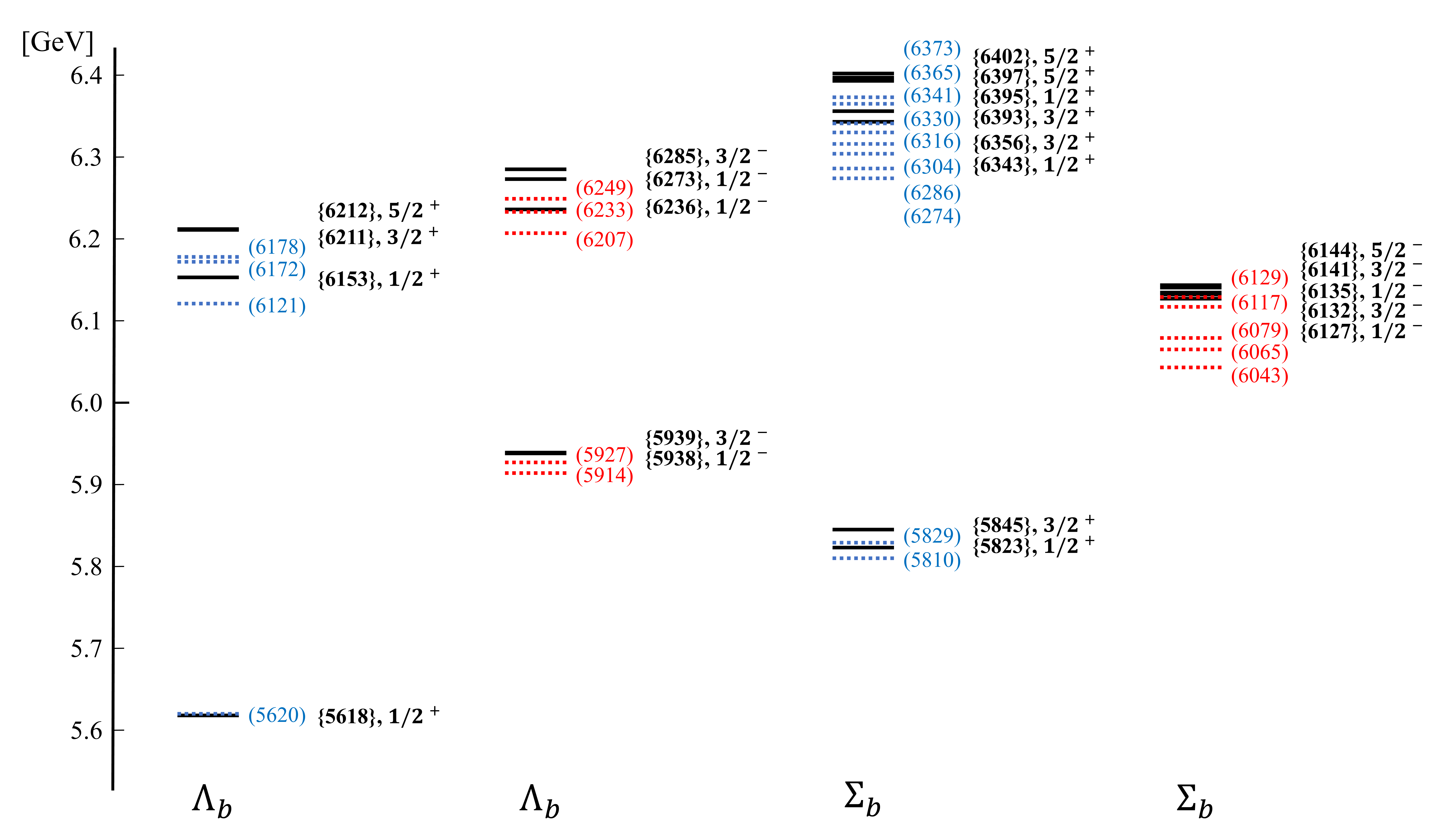}
  \caption{Energy spectra of $\Lambda_b$ and $\Sigma_b$ from our model with Y-potential (blue or red dotted lines), and from the three-quark model in Ref. \cite{yoshida} (black solid lines).}
\label{figurebottom3q}
\end{figure*}

Here, we compare the numerical results from our heavy-quark--diquark model and those from a constituent quark model containing two light quarks and one heavy quark, which we call the three-quark model.
In Ref.~\cite{yoshida}, $\Lambda_{c/b}$, $\Sigma_{c/b}$, and $\Omega_{c/b}$ baryons were analyzed, where the Schr$\ddot{\rm o}$dinger equation with three quarks was solved by the Gaussian expansion method \cite{GEM1, GEM2}.

In Figs. \ref{figurecharm3q} and \ref{figurebottom3q}, we show the comparison of the energy spectra of $\Lambda_{c/b}$ and $\Sigma_{c/b}$.
We can see that the masses from the three-quark model tend to be heavier than those from our model.
Also, in the excited states, the width of energy splitting from three-quark model is narrower than our model. 
For example, the value of splitting between the $\Lambda_c (1/2^-)$ and $\Lambda_c (3/2^-)$ is 36 MeV for our model, while Ref.~\cite{yoshida} obtained 2 MeV. 
This difference between two models is caused by the choice of potential parameters.
In Ref.~\cite{yoshida}, most of parameters were determined so as to reproduce the masses of hyperons (baryons containing three light quarks with at least one strange quark), while some parameters in our model are determined from the singly charmed or bottom baryons. 
As a result, the coefficient of the spin-orbit potential is $\eta=0.026$ in Ref.~\cite{yoshida}, while $\eta=0.2494$ in our model.
Since the strength of the spin-orbit potential is related to the mass splitting, thus the mass splitting by the spin-orbit potential in Ref.~\cite{yoshida} is much smaller than our results. 

Furthermore, due to the difference of the spin-orbit potential parameter $\eta$, the ordering of total angular momenta $J$ is different from that in our model. For example, in Fig.~\ref{figurepotsplit}, the ordering of the negative-parity $\Sigma_c$ from our model changes from $1/2^- = 3/2^- < 1/2^- = 3/2^- = 5/2^-$ to $1/2^- < 1/2^- < 3/2^- < 3/2^- <5/2^-$ adding the spin-orbit potential $V_{so}$. On the other hand, the ordering from Ref.~\cite{yoshida} is $1/2^- < 3/2^- < 1/2^- < 3/2^- < 5/2^-$ as shown in Fig.~\ref{figurecharm3q}, since the strength of spin-orbit potential is weak.

In conclusion, comparing with the experimental data in PDG \cite{PDG}, the results from our heavy-quark--diquark model are closer than those from the three-quark model in Ref.~\cite{yoshida}.

\section{Conclusion} \label{Sec_5}
In this paper, chiral properties and dynamics of the color-$\bar{\bf 3}$, spin-0 and -1 diquarks are studied.
The scalar (S, $0^+$) diquark and pseudoscalar (P, $0^-$) diquarks are paired into a chiral $(\bar{3},1)+(1,\bar{3})$ representation, while the axial-vector (A, $1^+$) and vector (V, $1^-$) diquarks are assigned to a chiral $(3,3)$ representation. The S, P, and V diquarks form anti-triplet ($\bar{\bf 3}$) representationss in flavor $SU(3)$, while the A diquark belongs to a $SU(3)$ sextet ({\bf 6}).

We have then constructed a chiral effective theory of the diquarks in the form of the linear-sigma model.
The masses of the S and P diquarks are given in term of three mass parameters, $m_{S0}^2$, $m_{S1}^2$ and $m_{S2}^2$. The first one $m_{S0}^2$ is chiral invariant, 
while the latter two generate the diquark masses associated with a symmetry breaking condensate 
of the meson field $\Sigma$.
Furthermore, the second one $m_{S1}^2$ represents the contribution of the axial $U(1)$ anomaly.
The mass splitting of the S and P diquarks are induced by the symmetry breaking terms.
In turn, if the chiral symmetry is fully restored, the S and P diquarks become degenerate, 
having the mass $m_{S0}$.

For the A and V diquarks, we also have three mass parameters, 
$m_{V0}^2$, $m_{V1}^2$ and $m_{V2}^2$, where the first one is chirally invariant 
and the other two are from the chiral symmetry breaking terms.
Similarly to the S and P diquarks, the mass difference between the A and V diquarks is induced 
by the $m_{V1}^2$ and $m_{V2}^2$ parameters.

We have obtained mass formulas for the S, P, A and V diquarks by further taking into account explicit flavor symmetry breaking due to a heavy mass of strangeness.
We find a few interesting relations among the diquark masses. 
One is the inverse hierarchy of the P diquarks, 
for which the diquark with a strange quark becomes lighter than the non-strange one, $M_{qq}(0^-) > M_{qs}(0^-)$.
The other is a generalized Gell-Mann--Okubo mass formula for the A diquarks, 
$[M_{ss}(1^+)]^2 - [M_{qs}(1^+)]^2 = [M_{qs}(1^+)]^2 - [M_{qq}(1^+)]^2$, where the 
squared diquark masses are equally spaced with the increased number of strange quarks.

In order to quantify the masses of the diquarks in a more realistic manner, 
we study the spectrum of singly heavy baryons using the heavy-quark--diquark model.
The heavy-quark--diquark model describes singly heavy baryons, such as $\Lambda_{c/b}$ and $\Sigma_{c/b}$, as two-body systems of a heavy quark and a diquark.
The potential between the two is given by
a linear confinement plus Coulomb potential with additional spin-dependent forces.

The parameters of the chiral effective Lagrangian and the heavy-quark--diquark potential are fixed
so as to reproduce the low-lying masses of the singly heavy baryons.
We have calculated the negative- and positive-parity excited states of the singly heavy baryons as well as
the ground states. The obtained spectra are compared with the current experimental data and also the 
standard three-quark descriptions in the constituent quark model.

The specific features of the diquark spectrum are also realized in the baryon spectrum.
The inverse mass hierarchy of the P diquarks causes the inverse mass hierarchy of singly heavy baryons in 
the $\rho$-mode excited state \cite{scalar}. 
The generalized Gell-Mann--Okubo mass formula is reflected in the spectrum of $\Sigma_Q, \Xi'_Q$ and $\Omega_Q$ baryons.

The energy spectra from our heavy-quark--diquark model are mostly consistent with the experimental data in the ground state. However, our results are lighter than the observed masses in the excited states.
In order to correctly reproduce these experimental data, we need to readjust the parameters of the central potentials.
 Also we have compared our results with the spectra from the three-quark model in Ref.~\cite{yoshida}. Here the energy splittings from our model are larger than those from the constituent three-quark model, and this difference is caused by the different strength of the spin-orbit potential. As a consequence, our model reproduced the results closer to the observed spectra. 
 
In this paper, we have introduced the well-known $S$-wave diquarks, S ($0^+$) and A ($1^+$), and their
chiral partners. By applying them to the singly heavy baryons, 
we find all the ground-state baryons are consistent
with the three-body quark models, while the $P$-wave excited states are not fully recovered.
A missing piece is the flavor $\bf 6$ vector diquarks. They should appear in the negative-parity baryons
with flavor $\bf 6$, {\it i.e.}, $\Sigma_Q$, $\Xi'_Q$ and $\Omega_Q$ excitations.
It is necessary to introduce a set of tensor diquarks with color $\bar{\bf 3}$ and flavor $\bf 6$ \cite{Shuryak:2003zi, Hong:2004xn, Shuryak:2005pk}.

Several future directions of the use of the diquark effective theory are quite interesting.
So far, we considered only color-$\bar{\bf 3}$ diquarks, while the other choice, the color-{\bf 6} diquark is
possible. It may not appear in the singly heavy baryon systems as it cannot make color singlet state with
a heavy quark. It, however, may appear in multi-quark systems, such as tetraquarks, $\bar Q\bar Q qq$ and pentaquarks, $\bar Q qq qq$ \cite{Chen:2016qju, Hosaka:2016pey, Chen:2017vko, Olsen:2017bmm, Ali:2018bdc, Liu:2019zoy}. 

It is also extremely interesting and important to reveal how chiral symmetry works on the diquarks at
finite temperature and baryon/quark density including the color superconducting phase \cite{Alford:1997zt, Rapp:1997zu}. In fact, we have found that in the course of chiral symmetry restoration,
the masses of the S and A diquarks may be inverted and the A diquarks may become dominant at
the vicinity of chiral phase transition.

\section*{Acknowledgments}

This work was supported in part by JSPS KAKENHI Grants No.~JP20K03959 (M.O.), No.~JP19H05159 (M.O.), No.~JP20K14476 (K.S.), No.~JP17K14277 (K.S.),~No. JP21J20048 (Y.K.), and also by NNSFC~(No. 11775132) (Y.R.L.).




\bibliography{ref-vdiquark}
\end{document}